\documentclass[12pt]{article}
\usepackage{epsfig}

\hoffset = - 0.6 truecm \voffset=-1.5 truecm
\def\lsim{\raise0.3ex\hbox{$<$\kern-0.75em\raise-1.1ex\hbox{$\sim$}}}
\def\gsim{\raise0.3ex\hbox{$>$\kern-0.75em\raise-1.1ex\hbox{$\sim$}}}

\begin{document}
\begin{center}
{\large\bf  Microscopic Models of Heavy Ion Interactions}
\vskip 1 truecm
{\bf A. Capella}

{\it Laboratoire de Physique Th\'eorique (UMR CNRS N$^{\circ}$
8627),}\\
  {\it Universit\'e Paris XI, B\^atiment 210, 91405 Orsay Cedex, France,}\\

{\it E-mail : alphonse.capella@th.u-psud.fr}
\end{center}
\vskip 2 truecm

\begin{abstract}
An introduction to dynamical microscopic models of hadronic and nuclear
interactions is presented. Special emphasis is put in the relation
between multiparticle production and total cross-section contributions.
In heavy ion collisions, some observables, considered as signals of the
production of a Quark Gluon Plasma (QGP), are studied. It is shown that
they can only be described if final state interactions are introduced.
It is argued that the cross-sections required are too small to drive
the system to thermal equilibrium within the duration time of the final
state interaction.
\end{abstract}
\vskip 2 truecm

  LPT Orsay 03-36 \par
May 2003

\vskip 2 truecm

\begin{center}
{\it Lectures at VIII Hispalensis International Summer School,
S\'eville (Spain),}\par
{\it  to be published by Springer Verlag}
\end{center}

\newpage
\pagestyle{plain}
\section{Introduction}
\label{sec:1}
Statistical QCD predicts the existence of new states of matter at high
temperature $T$ and high baryon number densities $\mu$ (baryochemical
potential). The phase diagram is schematically represented in Fig.~1.
Let us discuss first the phase transition at high $T$ and small $\mu$
(i.e. when baryon and antibaryon densities are approximately equal). At
$\mu = 0$, lattice calculations \cite{1r} show a phase transition to a
deconfined plasma of quarks and gluons (QGP) at a critical temperature
$T_c \sim 150 \div 200$~MeV, corresponding to an energy density
$\varepsilon$ -- an order of magnitude higher than that of ordinary
nuclear matter ($\varepsilon_0 = 170$~MeV/fm$^3$). The results are
shown in Fig.~2. Below the critical temperature $T_c$, we heat the
system and the temperature increases towards $T_c$. At $T \sim T_c$, the
temperature remains constant (the energy given to the system is used to
increase its latent heat) and the energy density $\varepsilon$
increases sharply. For $T > T_c$ the temperature increases again with
$\varepsilon/T^4$ approximately constant, as it should be for an ideal
gas (Stefan-Boltzmann limit). The sharp increase of $\varepsilon/T^4$
near $T_c$ is due to the increase in the number of degrees of freedom
of the system from a hadronic phase ($T < T_c$) to a plasma of quarks
and gluons ($T > T_c$). \par

In the ideal case of pure gauge QCD (the so-called ``quenched''
approximation in which dynamical quarks are absent) the phase
transition is first order. In the presence of dynamical quarks the
situation is more complicated and the order of the phase transition
depends on the number of flavors. A restoration of chiral symmetry also
takes place at the same $T_c$. \par

The region $\mu \sim 0$ studied in lattice QCD corresponds to the
conditions in high-energy heavy ion collisions at mid-rapidities. It
also corresponds to the conditions in the primordial universe. In the
opposite conditions, i.e. low temperature and high baryon densities
(i.e. the conditions at the center of neutron stars) a new phase (or
phases) is expected \cite{2r} producing a color superconductor in which
pairs of quarks condensate -- in a way similar to electron (Cooper)
pairs in QED. \par

A possibility to create in the laboratory the high energy densities and
temperatures required for the production of QGP is via head-on heavy
ion collisions at high energy. In $AA$ collisions, the energy density
can be evaluated using the Bjorken formula

$$\varepsilon \sim \left ( {dN \over dy}\right )_{y^* \sim 0}^{hadrons}
\ {<m_T> \over \tau_0 \pi R_A^2} \ .$$

\noindent Here $dN/dy$ is the number of produced hadrons per unit
rapidity and $<m_T>$ is their average energy. $R_A$ is the nuclear
radius and $\tau_0$ an average formation time. Taking the customary
value $\tau_0 \sim 1$~fm, we obtain for central $Au$ $Au$ collisions at
RHIC an energy density $\varepsilon \sim 4$~GeV/fm$^3$, which
appears to be high enough for QGP production. \par

Does it mean that QGP production in high-energy heavy ion collisions
follows from QCD~? The answer is negative. Indeed, QGP formation is
predicted by statistical QCD, i.e. QCD applied to a system in thermal
equilibrium. Therefore, one of the main issues in heavy ion physics is
to determine whether the produced final state reaches thermal
equilibrium. This depends, of course, on the strength and time duration
of the final state interaction and can only be decided with the help of
experiment. The observables which provide the most reliable information
are the so-called signals of QGP. In this lecture I will discuss two of
the most important signals~: the particle abundance (in particular,
hyperon and antihyperon production) and the $J/\psi$ suppression. The
latter could be due to the deconfinement or ``melting'' of the
$c\bar{c}$ bound state in the plasma \cite{3r}. However, a similar
phenomenon is observed in $pA$ collisions, where the produced densities
are too small for QGP formation. As for the former, an argument in
favor of equilibrium is the fact that particle abundances are well
described using statistical models \cite{4r}. Moreover, these models
provide a natural explanation \cite{5r} of the increase of the relative
yields of strange particles in central nucleus-nucleus as compared to
$pp$ collisions (strangeness enhancement). However, one should take
into account that these models are also very successful \cite{6r} in
$pp$ and even in $e^+e^-$ interactions, where QGP is not produced. \par

Therefore, it is important to study these observables in the framework
of microscopic models which are successful in describing $pp$ and $pA$
interactions and can be generalized to heavy ion collisions.
Models of this type [7-10] are called string models, in which
particle production takes place in the form of strings (chromo-electric
flux tubes) stretched between constituents of complementary color
charge. In QCD, the force between complementary color charges (such as
a quark and an antiquark of same color), are small at distances
smaller than the hadron radius. However, at larger distances, the
potential of the chromo-electric field increases linearly with the
distance (confinement). Due to the strong force generated by this
potential it is not possible to isolate the two color charges. Indeed,
in the attempt to separate them, the potential energy of the system
increases very rapidly and is converted into mass via the creation of
quark-antiquark pairs. This results in the production of hadrons
(mostly mesons), formed by recombination of a quark and an antiquark of
adjacent pairs. The lines of force of the chromo-electric field are
strongly colimated along the axis determined by the two color charges.
Hence the string-like or jet-like shape of the set of produced
hadrons.\par

As a starting point, one usualy assumes in these models that particles
produced in different strings are independent. In this case thermal
equilibrium cannot be reached, no matter how large the energy density
is. Indeed, in this case a large energy-density is the result of piling
up a large number of independent strings. In other words, some
``cross-talk'' between different strings is needed in order to
thermalize the system. \par

The assumption of independence of strings works remarkably well in $hh$
and $hA$ interactions [7-10] -- even in the case of event samples with 5 or
6 times the average multiplicity -- indicating that no sizable final
state interaction is present in these reactions. However, it is clear
that in heavy ion collisions, where several strings occupy a transverse
area of 1 fm$^2$, the assumption of string independence has to break
down. This is indeed the case. As we shall see, some data cannot be
described without final state interaction. It could have happened that
this final state interaction is so strong that the string picture
breaks down and becomes totally useless. This does not seem to be the
case. On the contrary, present data can be described using the particle
densities computed in the model as initial conditions in the gain and
loss (transport) equations governing the final state interaction. The
interaction cross-section turns out to be small (a few
tenths of a mb). Due to this smallness and to the limited interaction
time available, final state interaction has an important effect only on
rare processes, in particular $\Xi$, $\Omega$ and $J/\psi$ production,
or particle yields at large $p_T$. The bulk of the final state is not
affected. \par

The plan of these lectures is as follows~: in Section \ref{sec:2}, I
introduce the general framework of high-energy scattering in a hadronic
language with special emphasis on the unitarity condition and its
implementation in eikonal and Glauber Models. This section contains
also some rudiments of Regge poles and the concept of Pomeron. In
Section \ref{sec:3}, I introduce a microscopic model~: the Dual Parton
Model (DPM) and compute the charged particle multiplicities as a
function of energy and centrality (the latter characterizes the
impact parameter
of the collision in a way which is experimentally measurable). In
Section \ref{sec:4}, I study the so-called stopping power, i.e. the
fate of the nucleons of the colliding nuclei.
Particle abundances are studied in Section \ref{sec:5} and $J/\psi$
suppression in Section \ref{sec:6}. Section \ref{sec:7} contains the
conclusions.

\section{High-Energy Scattering~: Hadronic Picture}
\label{sec:2}

\subsection{General Framework} \label{subsec1}

An important property of strong interactions is the unitarity of the
$S$ matrix operator

\begin{equation} \label{1e} SS^+ = S^+S = 1 \ . \end{equation}

\noindent The $S$ matrix is the operator that transforms free states at
time $t = - \infty$ into free states at times $t = + \infty$. If we
write

\begin{equation} \label{2e} S = 1 + i T \ , \end{equation}

\noindent where $<a|T|b>$ is the transition amplitude between free
states $a$ and $b$, the unitarity condition (\ref{1e}) reads

\begin{equation} \label{3e} 2\ Im \ <a|T|b> = \sum_{\hbox{all $n$}}
<a|T|n> \ <n|T|b>^* \ . \end{equation}

\noindent Here $n$ denotes any state that can be reached from both $a$
and $b$. For $a$ identical to $b$, eq. (\ref{3e}) reduces to

\begin{equation} \label{4e} 2\ Im \ <a|T|a> = \sum_{\hbox{all $n$}}
\left ( <a|T|n> \right )^2 \ . \end{equation}

\noindent A two body amplitude $1 + 2 \to 3 + 4$ depends on two
independent variables~:

\begin{equation} \label{5e} s = \left ( p_1 + p_2 \right )^2 \quad ,
\quad t = \left (p_1 - p_3\right )^2 \quad , \quad u = \left (p_1 -
p_4\right )^2 \ , \end{equation}

\noindent with $s + t + u = m_1^2 + m_2^2 + m_3^2 + m_4^2$. When $a$
and $b$ are identical, $t = 0$. Since the sum in the r.h.s. of
(\ref{4e}) builds up
the total $1 + 2$ cross-sections, we obtain from (\ref{4e}) the optical
theorem, which relates the total cross-section to the forward ($t = 0$)
imaginary part of the elastic amplitude $T_{el}(s,t)$

\begin{equation} \label{6e} \sigma_{tot}(s) = {4 \pi \over s} \ Im\
T_{el}(s, t=0) \ . \end{equation}

Let us now examine some of the implications of the unitarity condition
(\ref{4e}). It is convenient to separate the states $n$ into two
classes. The first class consists of all inelastic states containing no
large rapidity gap between the produced particles. The second class
consists of those events with at least one large rapidity gap.
Obviously, elastic scattering produces a state of the second class. We
know experimentally that the first class of events gives the most
important contribution to the total cross-section. It is convenient to
depict such a contribution in the form of a diagram as shown in Fig.~3a.
(Its interpretation will be discussed below). The important point is
that the final state in Fig.~3a has also to be included as intermediate
state in 4. This is precisely the contribution of the elastic state
which, as discussed above, belongs to the second class of intermediate
states. One obtains in this way the diagram shown in Fig.~3b. Here the
lines with a cross denote the contribution of the on-shell initial
state to the elastic amplitude. Repeating such an iteration, one is led
to the multiple scattering diagram depicted in Fig.~3c. One has to add
up all the diagrams in Figs.~3. In fact, inelastic states (for instance
$pN^*$ and $N^*N^*$ in the case of $pp$ scattering) also contribute as
intermediate states, so that the vertex function (blob) of the diagrams
in Fig. 3b and 3c can be quite complicated. There are also
contributions due to intermediate states $n$ containing more than one
rapidity gap. However, their contributions are small at present
energies and will not be considered here. \par

The interpretation of the diagram in Fig.~3a originates from the claim
that the inelastic intermediate states (without rapidity gap) generate
a Regge pole called the Pomeron (see Section \ref{subsec5}). All
diagrams can be treated as Feynman diagrams in a field theory called
Reggeon Field Theory or Gribov's Reggeon Calculus \cite{11r}. When only
the diagrams in Figs.~3 are kept, one obtains, as a particular case, the
so-called eikonal model for hadron-hadron scatttering or the Glauber
model in interactions involving nuclei.

\subsection{The Eikonal and Glauber Models} \label{subsec2} In the case
of hadron-hadron interactions the diagram in Fig.~3a (Pomeron) is also
called the Born term. Its contribution to the total cross-section is
parametrized in the form $g_{13} (t) g_{24}(t) s^{\Delta}$ where $g$
are coupling constants and $\Delta$ a parameter (see Section
\ref{subsec5}). Let us consider the case of $pp$ scattering  and let us
assume for simplicity an exponential dependence, in $t$, i.e.

\begin{equation} \label{7e} g_{13}(t) = g_{24}(t) = A \exp (Bt) \ .
\end{equation}

\noindent In this case, all loop integrals in Figs.~3b and 3c can be
performed analytically. At a given $s$ there are only two free
parameters $A^2s^{\Delta}$ and
$B$, which can be determined from the experimental values of
$\sigma_{tot}$ and $\sigma_{el}$ (or from $\sigma_{tot}$ and the slope
of the elastic amplitude). For $s$ variable there is a third parameter
$\Delta$, and the values of $\sigma_{tot}$ and $\sigma_{el}$ at several
energies can be described in this way.\par

A contribution to the total cross-section in $s^{\Delta}$ with $\Delta
> 0$ violates the Froissard bound \cite{12r} (consequence of
unitarity). Note that $\Delta > 0$ is needed in order to describe the
rise with energy of $\sigma_{tot}$. However, when all multiple
scattering terms are added, one obtains, at high energy, $\sigma_{tot}
\propto \ell n^2s$, i.e. the maximal increase with $s$ allowed by the
Froissard bound. \par

It is convenient to work in the impact parameter $(b)$ representation.
The scattering amplitude in $b$-space is defined as $T(s,b) = (1/2 \pi
) \int d^2 q_T \exp (- i \vec{q}_T\cdot \vec{b})T(s,t)$ where $t = - q_T^2$.
Using this transformation it is easy to see that at fixed $b$ the
contribution of a diagram involving $n$ exchanges (Fig.~3c) is just the
$n$-th power of the Born term (Fig.~3a) times some trivial
combinatorial factors. Furthermore the sum over $n$ has a simple
expression (see below, footnote 1). All these features will now be
described in detail in the
case of the Glauber model. \par

Let us consider, for definiteness, a proton-nucleus ($pA$) scattering.
In this case there are no free parameters involved. Indeed, the
$t$-dependence (or $b$-dependence) of the proton vertex function can be
neglected in comparison with the fast variation of the nuclear one.
The latter is known from the nuclear density. Moreover, in this case
the Born term at $t = 0$ is just the inelastic proton-nucleon
cross-section, which is known experimentally. The main formula of the
probabilistic Glauber model is the one that gives the cross-section
$\sigma_n$ for $n$ inelastic collisions of the projectile with $n$
nucleons of the target nucleus (the remaining $A - n$ nucleon which do
not participate in the interaction are called spectators), at fixed
impact parameter $b$~:

\begin{equation}
\label{8e}
\sigma_n (b) = {A \choose n} \left ( \sigma_{inel} \ T_A(b) \right
)^n \left ( 1 - \sigma_{inel} \ T_A(b) \right )^{A-n} \ .
\end{equation}

\noindent Here $\sigma_{inel}$ is the proton-nucleon inelastic
cross-section and $T_A(b)$ is the nuclear profile function (obtained by
integrating the nuclear density~: $T_A(b) = \int_{-\infty}^{+\infty} dZ
\rho_A(Z,b)$~; $\int d^2b T_A(b) = 1$). Equation (\ref{8e}) is just the
Bernoulli's formula for composite probabilities. The first factor is a
trivial combinatorial factor corresponding to the different ways of
choosing $n$ nucleons out of $A$. The second one gives the probability
of having $n$ inelastic $pN$ collisions at given $b$. The third one is
the probability that the remaining $A - n$ nucleons do not interact
inelastically. Let us consider first a term with two collisions both of
which are inelastic. The corresponding cross-section is
$\sigma_{2}^2(b) = {A \choose 2} (\sigma_{inel} T_A(b))^2$ i.e. a
positive term. Let us now consider the case of two collisions only one
of which is inelastic. The corresponding (interference) term is
$\sigma_2^1(b)$ obtained from Eq. (\ref{8e}) by putting $n = 1$ and taking
the second term in the expansion of the last factor. We get
$\sigma_2^1(b) = - A(A-1)(\sigma_{inel}T_A(b))^2$. We see that
$\sigma_2^1(b) = - 2 \sigma_2^2(b)$. Thus, a rescattering term
containing two collisions gives a negative contribution to the total
$pA$ cross-section.
  \par

Let us now consider their contributions to $d\sigma /dy$. They are
given by $\sigma_2^1(b) + 2 \sigma_2^2(b) = 0$. Indeed, in the case of
a double inelastic collision, the triggered particle can be emitted in
either of them -- hence an extra factor 2. We see, in this way that the
different contributions to $d\sigma/dy$ of a double scattering diagram
cancel. It is easy to see that this cancellation is valid order by
order in the total number of collisions. This can also be seen as
follows. The total inelastic cross-section for $pA$ collision in the
Glauber model is given by the well known expression\footnote{In the
literature, the optical limit expression is often used instead of Eq.
(\ref{8e}), namely $$\sigma_n(b)= \left ( \sigma_{inel} A T_A(b)\right
)^n \exp \left ( - \sigma_{inel} A T_A(b) \right )/ n!\ .$$ In this
case we get $\sigma_{inel}^{pA}(b) = 1 - \exp (-\sigma_{inel} A
T_A(b))$, which coincides with Eq. (\ref{9e}) in the large $A$ limit.
The same type of expressions are obtained in the eikonal model, since,
in this case, the number of rescatterings is infinite.}

\begin{equation}
\label{9e}
\sigma_{inel}^{pA} (b) = \sum_{n=1}^A \sigma_n (b) = 1 - \left ( 1 -
\sigma_{inel} \
T_A(b) \right )^A \ . \end{equation}

\noindent This expression contains a term in $A^1$ (single-scattering
or Born term). It
also contains contribution from multiple scattering with alternate
signs (shadowing corrections). Numerically,
it behaves as $A^{\alpha}$ with $\alpha \sim 2/3$. The single
particle inclusive
cross-section is given by

\begin{equation}
\label{10e}
{d\sigma^{pA} \over dy}(b) \propto \sum_{n=1}^A n\ \sigma_n(b) = A \
\sigma_{inel}\
T_A(b) \ . \end{equation}

\noindent We see that here multiple-scattering contributions cancel
identically and only the Born term is left (impulse approximation).
As a consequence of this
cancellation the $A$-dependence of $d\sigma/dy$ in $pA$
interactions behaves as $A^1$. In the case of $AB$ collisions it
behaves as $AB$ and $dN^{AB}/dy = (1/\sigma_{AB})
d\sigma^{AB}/dy$ is proportional to $A^{4/3}$, i.e. to the number of
binary collisions --
rather than to the number of participants, as one would naively
expect (see Section \ref{subsec3.2}).

\subsection{Shadowing Corrections in the Inclusive Cross-Section}
\label{subsec3} The interest of this way of looking at the Glauber
model resides in the fact that, as discussed in Section \ref{subsec2},
it provides strict relations between contributions to the total
cross-section and contributions to various inelastic processes of
multiparticle production, which make up the total cross-section via
unitarity. These relations are the so-called Abramovsky-Gribov-Kancheli
(AGK) cutting rules \cite{13r}, and have a general validity in RFT. The
absence of shadowing corrections in the inclusive cross-section, Eq.
(\ref{10e}), is called the AGK cancellation. As mentionned above, in
the eikonal and Glauber models only the initial state is present in the vertex
function (blob). Thus a secondary can only be produced in an
interaction and the AGK cancellation is exact. In a general theory with
a more complicated vertex function, the triggered particle can be
produced in the blob. This gives rise to a violation of the AGK
cancellation -- which is responsible for the shadowing corrections to
the inclusive spectra. Indeed, if the measured particle (trigger) is
produced in the vertex function of a double scattering diagram, the
extra factor 2 in $\sigma_2^2(b)$ is not present and the AGK
cancellation is not valid. In this case the shadowing corrections in
the inclusive cross-section are the same as in the total cross-section.
This is the physical origin of the AGK violations present in the
microscopic model described in Section \ref{sec:3}. It is clear that if
the blob has a small extension in rapidity, production from the blob
will mainly contribute to the fragmentation region. Therefore, at
mid-rapidities, and large energy, the AGK cancellation will be valid. \par

Let us consider next the contribution to the total cross-section
resulting from the diffractive production of large mass states.
Clearly, this is equivalent to an increase of the rapidity extension of
the blob -- which, in this case, can cover the mid-rapidity region.
Therefore, shadowing corrections to the single particle inclusive
cross-section can be present, at mid-rapidities, provided the measured
particle is part of the diffractively produced system. We see in this
way that shadowing corrections to $d\sigma/dy$ are related to
diffractive production of large mass systems. The theoretical
expression of the diffractive cross-section is well-known. It has also
been measured experimentally and, thus, the shadowing corrections can
be computed with no free parameters. In these lectures I will not
elaborate any further on this last point. More details, as well as
numerical calculations, can be found in \cite{14r}.

\subsection{Space-Time Development of the Interaction : Absence of
Intra-Nuclear Cascade}
\label{subsec4}
The reggeon calculus or reggeon field theory (RFT) \cite{11r} provides
the field theoretical formulation of the eikonal (for $hh$ collisions)
or the Glauber (for $hA$ and $AB$) models, valid at high energies. The
main difference between the RFT and the Glauber model is that, at high
energies, the coherence length is large and the whole nucleus is
involved in the interaction. Moreover, due to the space-time
development of the interaction, when, at high energy, a projectile
interacts inelastically with a nucleon of the nucleus, the formation
time of (most of) the produced particles is larger than the nuclear
size and, thus, most particles are produced outside the nucleus. Only slow
particles in the lab reference frame are produced inside the nucleus
and can interact with the nucleons of the nucleus they meet in their
path (intra-nuclear cascade). At high
energies, most of the produced
particles have left the nucleus at the time they are formed. This near
absence of nuclear cascade is well known experimentally. Actually, it
constituted for a long time one of the main puzzles of high energy
hadron-nuclear interactions. \par

Another consequence of the space-time development of the interaction,
is that planar multiple-scattering diagrams give a vanishing
contribution to the total cross-section at high energies. Indeed, as
discussed above, the formation time of the multiparticle state is
larger than the nuclear size and, therefore, there is no time for its
rescattering with other nucleons of the nucleus. The relevant multiple
scattering diagrams are non-planar ones, describing the ``parallel''
interactions of different constituents of the projectile with the
target nucleons
(in the case of an $hA$ collision). This picture is in clear contrast
with the Glauber model, in which the projectile undergoes successive
(billiard ball type of) collisions with the nucleons of the target.
\par

In spite of these differences, one recovers the Glauber formula in
first approximation. As discussed above, this formula corresponds to
the contribution of
the initial state (on-shell projectile pole) to the various
rescattering terms. In RFT one has, besides these contributions, also
the contributions due to low mass and high mass diffractive excitations
of the projectile. The latter are very important since, as we have seen
in Section \ref{subsec3}, they give rise to shadowing corrections to
the inclusive cross-section.\par

\subsection{Regge Poles : The Pomeron} \label{subsec5} Regge poles
\cite{15r} \cite{16r} play an important role in high-energy interactions. It is
important to give some basic concepts on Regge poles. Indeed, they
provide a theoretical basis for the parametrization of the Born term in
multiple-scattering models. More precisely, they give an important
connection between high-energy behaviour and the spectrum of resonances
in the $t$-channel. Also, it is very important for our purpose in these
lectures that they allow to determine, to a large extent, the
momentum distribution functions of partons in hadrons, as well as their
fragmentation functions. These are the main ingredients of the
microscopic models introduced in the next section. \par

At high energies the exchange of a particle of spin $J$ and mass $M_J$
in the $t$-channel of a two-body amplitude $1 + 2 \to 3 + 4$ has the form

\begin{equation} \label{11e} T(s,t) = g_{13} \ g_{24} \ s^J/(M_J^2 - t)
\ . \end{equation}

\noindent When $J$ is large ($J > 1$), the cross-section increases
as a power $s^{J-1}$ and violates the Froissart bound. This problem
provided one of the main motivations for the introduction of Regge
poles. Note that Eq. (\ref{11e}) is valid only close to the pole, i.e.
for $t \sim M_J^2$. However, the physical values of $t$ for the
$s$-channel amplitude $1 + 2 \to 3 + 4$ are $t \ \lsim\ 0$. In the
Regge pole model, the behaviour (\ref{11e}) is modified for physical
values of $t$, as follows

\begin{equation} \label{12e}
T(s,t) = g_{13}(t) \ g_{24} (t) \ s^{\alpha (t)} \ \eta (\alpha (t)) \ .
\end{equation}

\noindent Here $g(t)$ are called Regge residues and $\alpha
(t)$ is a function, called Regge trajectory, such that $\alpha (t =
M_J^2) = J$. Thus, if $\alpha (t) \leq 1$ for physical values of $t$,
the Froissard bound will be satisfied, irrespective of the value of
$J$. For practical purposes one proceeds as follows. Let us consider the
existing particles and/or resonances with the conserved quantum numbers
(isospin, parity etc) of the $t$-channel. For each of them, let us plot
its spin versus its mass (see Fig.~4). This is the so-called
Chew-Frautschi plot \cite{16r}. (Actually, in a relativistic theory it
is necessary to consider separately even and odd values of spin). It
turns out that a large family of resonances, including $\rho$,
$\omega$, $f$ and $A_2$ lie practically in a single straight line
$\alpha_R (t) = \alpha_R(0) + \alpha't$. Other resonances as $K^*$ and
$\phi$ lie on parallel Regge trajectories with $\alpha (t) < \alpha_R
(t)$. Extrapolating to the physical region
$t \leq 0$ one finds an intercept $\alpha_R(0) \sim 1/2$ and a slope
$\alpha '_R \sim 0.9$~GeV$^{-2}$. Since $\alpha_R(t) \ \lsim\ 1/2$ for
$t \ \lsim\ 0$, the Froissard bound is respected. The function $\eta
(\alpha (t)) = - [1 + \sigma \exp (-i\pi \alpha (t))]/\sin \pi \alpha
(t)$ with $\sigma = + 1$ $(-1)$ for $J$ even (odd). $\sigma$ is called
the signature~: positive (negative) for $\sigma = + 1$ $(-1)$. Note
that for a trajectory of positive (negative) signature, the numerator
of $\eta$ cancels the poles in the denominator corresponding to odd
(even) values of spin. Note also that $\eta (\alpha (t))$ determines
the phase of the amplitude in terms of $\alpha (t)$. In this way the
phase of the amplitude is related to its high energy behaviour -- also
determined by $\alpha (t)$. (The validity of this relationship is much
more general than the Regge pole model). \par

The Regge pole model has many features that have been tested by
experiment (and no contradiction with experiment has been found). Among
its successes is the factorization of Regge residues, Eq. (\ref{12e}).
Another success is that, for any given two body process, when
resonances exist in its $t$-channel, a (shrinking) peak is found in the
$s$-channel amplitude near $t = 0$ -- as obtained from Eq.
(\ref{12e})\footnote{Assuming an exponential $t$-dependence for
$g(t)$, we obtain from Eqs. (\ref{7e}) and
(\ref{12e}), $T(s,t) \propto \exp [(2B+ \alpha ' \ell n\ s)t]$. We see
that the forward peak has a width that increases logarithmically with
increasing energy. This is called the shrinking of the forward peak and
has been observed experimentally. In this way $r \propto \ell n\ s$,
i.e. the effective radius of the hadron increases like $\ell n\ s$ as
$s \to \infty$}. On the contrary, if the $t$-channel is ``exotic''
(i.e. no known resonance exists having its quantum numbers), the
forward peak in the $s$-channel is absent. There is no exception to
this rule.\par

Unfortunately, there is an important caveat. All Regge trajectories
corresponding to known resonances have an intercept $\alpha (0) < 1$.
Therefore, it is not possible to explain the increase with energy
observed in total cross-sections. In order to do so, it is necessary to
postulate the existence of a Regge pole with vacuum quantum numbers
(so that it can be exchanged in all elastic amplitudes) and intercept
larger than one. It is called the Pomeron and is believed
to correspond to the exchange of glue-balls in the $t$-channel. To
compute its trajectory in QCD is a very difficult task. Both
non-perturbative and perturbative methods give indications that its
trajectory is slightly above one ($\alpha_P(0) \sim 1.1 \div 1.3$) as
required by experiment. In this way it is possible to explain the
increase with energy of total cross-sections, i.e. $\sigma_{tot} \propto
s^{\Delta}$ with $\Delta = \alpha_P(0) - 1 > 0$. \par

Note that for $\alpha_P(0) = 1$, the amplitude at $t = 0$ is purely imaginary.
With $\alpha_P(0)$ slightly above unity the ratio of real to imaginary
parts is small. Unfortunately, there is still a caveat. Since $\Delta =
\alpha_P(0) - 1 > 0$, the Pomeron exchange again violates the Froissard
bound. (This violation was one of the main motivations for Regge
models, as discussed at the beginning of this subsection). The only way
out is to use the Pomeron contribution as a Born term in a
unitarization scheme -- such as the eikonal or Glauber models discussed
above. Indeed, if the contribution of the Born term or Pomeron
(Fig.~1a) to $\sigma_{tot}$ behaves as $s^{\Delta}$, the sum of the
series (Figs.~1b and 1c) behaves as $s^{\Delta '}$ with $\Delta ' <
\Delta$. Moreover, in the limit $s \to \infty$ the power behaviour of
the Born term is converted into $(\ell n\ s)^2$, i.e. the maximal
energy growth allowed by the Froissard bound\footnote{This can be seen
as follows. If the cross-section of the Born term tends to infinity,
the inelastic cross-section of the eikonal sum, at fixed $b$, tends to
1 (see footnote 1). Upon integration in $b$ one obtains a geometrical
cross-section, proportional to $r^2 \propto (\ell n\ s)^2$ (see
footnote 2).}. \par

Technically, Regge poles are isolated poles in the complex angular
momentum plane. Their $s$-channel iteration (as in the eikonal model)
gives rise to cuts in this plane (Regge cuts). The latter violate
factorization. However, it turns out that the sum of all eikonal
diagrams is approximately factorizable.

\section{Microscopic String Models}
\label{sec:3}

\subsection{Hadron-Hadron Interactions}
\label{subsec3.1}
The Dual Parton Model (DPM) \cite{7r} \cite{8r} and the Quark Gluon
String Model
(QGSM) \cite{9r} \cite{10r} are closely related dynamical models of
soft hadronic
interactions. They are based on the large-$N$ expansion of
non-perturbative QCD\footnote{The Feynman graphs of a gauge field
theory with $N$ degrees of freedom can be classified according to their
topology. The graphs with the simplest topology are dominant. The
contribution of graphs with more complicated topology (characterized by
well defined topological indices) are suppressed by powers
of $1/N$ [17-19].} and on Gribov's Reggeon Field Theory (RFT)
\cite{11r}. Their main aim is to determine the mechanism of
multiparticle production in hadronic and nuclear interactions. The
basic mechanism is well known in $e^+e^-$ annihilation (Fig. 5). Here
the $e^+e^-$ converts into a virtual photon, which decays into a
$q\overline{q}$ pair. In the rest system of the virtual photon the
quark (colour 3) and the antiquark (colour $\overline{3}$)
separate from each other producing one string (or chain) of hadrons,
i.e. two back-to-back jets. Processes of this type are called
one-string processes. \par

In hadron-hadron interactions, a one-string mechanism is also possible but only
in some cases, namely when the projectile contains an antiquark (quark) of the
same type than a quark (antiquark) of the target, which can annihilate with
each other in their interaction. For instance in $\pi^+p$, the $\overline{d}$
of $\pi^+$ can annihilate with the $d$ of $p$ and a single string is stretched
between the $u$ of $\pi^+$ (colour 3) and a diquark $uu$ of $p$ (colour
$\overline{3}$). This mechanism is also possible in $\overline{p}p$
interactions (Fig. 6) but not in $pp$. This already indicates that it cannot
give the dominant contribution at high energy. Indeed, when taking
the square of the diagram
of Fig. 6 (in the sense of unitarity) we obtain a planar graph, which is the
dominant one according to the large-$N$ expansion. However, this only means
that this graph has the strongest coupling. Since flavour quantum numbers are
exchanged between projectile and target, this graph gives a contribution to the
total cross-section that decreases as an inverse power of $s$ ($1/\sqrt{s})$. A
decrease with $s$ is always associated with flavor exchange. For instance, the
charge exchange $\pi^-p \to \pi^0n$ cross-section also decreases as
$1/\sqrt{s}$. As we have discussed in Section \ref{subsec5}, only an
exchange in the $t$-channel with vacuum quantum numbers (Pomeron),
gives a contribution to $\sigma_{tot}$ which does not vanish
asymptotically. Actually, the diagram in Fig.~6 corresponds to the
exchange of a Reggeon, with intercept $\alpha_R (0) \sim 1/2$. \par

In order to prevent the exchange of flavour between projectile and
target, the $\overline{d}$ and $d$ have to stay, respectively, in the
projectile and target hemispheres. Since they are coloured, they must
hadronize stretching a second string of type $\overline{d}$-$d$. We
obtain in this way a two-string diagram (Figs. 7-9)). Taking the
square of this diagram, we obtain a graph with the topology of a
cylinder (Fig. 10). It turns out that this is the simplest topology one
can construct which does not vanish as $s\to \infty$ due to flavour
exchange. Therefore, we obtain in this way the dominant graph for
hadron-hadron scattering at high energy. The diagram in Fig. 10
corresponds to a
Pomeron (P) exchange and the graphs in Figs. 7-9 are called a cut Pomeron. Its
order in the large-$N$ expansion is $1/N^2$ [17-19]. Note that due to
energy conservation the longitudinal momentum fractions taken by the
two components at the string ends have to add up to unity.\par

There are also higher order diagrams (in the sense of the large-$N$
expansion) with 4, 6, 8 strings which give non-vanishing contributions
at high energy. An example of the next-to-leading graph for $pp$
interactions is shown in Fig. 9. It contains four strings -- the two
extra strings are stretched between sea quarks and antiquarks. The
square of this graph corresponds \cite{20r} to a two Pomeron exchange
(Fig.~1b) and
has the topology of a cylinder with a handle. Its order in the
large-$N$ expansion is $1/N^4$. The one with six strings corresponds
\cite{20r} to a three Pomeron exchange (Fig.~1c) and to the topology of
a cylinder with two handles (order $1/N^6$), etc. \par

In this way, the large-$N$ expansion provides the microscopic
(partonic) description of the Reggeon and Pomeron exchanges and of
their $s$-channel iterations (Figs.~3), which were discussed in Section
\ref{sec:2} in a hadronic picture. \par

The single particle inclusive spectrum is then given by \cite{7r}

\begin{eqnarray}
\label{13e} &&{dN^{pp} \over dy}(y) = {1 \over \sum\limits_n \sigma_n} \sum_n
\sigma_n \left ( N_n^{qq-q_v}(y) + N_n^{q_v-qq}(y) + (2n-2)
N_n^{q_s-\overline{q}_s}(y) \right ) \nonumber \\ &&\simeq N_k^{qq-q_v}(y) +
N_k^{q_v-qq}(y) + (2k-2) N_k^{q_s-\overline{q}_s}(y) \end{eqnarray}

\noindent where $k = \sum\limits_n n \sigma_n/\sum\limits_n \sigma_n$ is the
average number of inelastic collisions. Note that each term consists
of $2n$ strings,
i.e. two strings per inelastic collisions. Two of these strings, of
type $qq$-$q$, contain the diquarks of the colliding protons. All other
strings are of type $q$-$\overline{q}$.\par

The weights $\sigma_n$ of the different graphs, i.e. their contribution
to the total cross-section, cannot be computed in the large-$N$
expansion. However, as discussed above there is a
one-to-one correspondence \cite{20r} between the graphs in the large-$N$
expansion and those in a multiple scattering model (Figs.~1). Thus, we
use the weights obtained from the latter -- with the parameters
determined from a fit to total and elastic cross-sections (see
Section \ref{sec:2}). At SPS energies we get $k = 1.4$ and at RHIC
$k=2$ at
$\sqrt{s} = 130$~GeV and $k = 2.2$ at $\sqrt{s} = 200$~GeV
\cite{21r}.\par

The hadronic spectra of the individual strings $N(y)$ are obtained from
convolutions of momentum distribution functions, giving the probability to
find a given constituent (valence quark, sea quark of diquark) in the
projectile or in the target, with the corresponding fragmentation
functions. Let us consider,
for instance, one of the two $qq$-$q$ strings in Fig.~9. As shown in
this figure, the
total energy $\sqrt{s}$ in the $pp$ center of mass frame (CM) is shared
between the two strings. If $\sqrt{s}_{str}$ denotes the invariant mass
of a string, we have $s_{str} = s x_2(1-x_1)$ where $1 - x_1 = x_+$ and
$x_2 = x_-$ are the light-cone momentum fractions of the constituents
at the string ends, $qq$ and $q$, respectively. For massless quarks,
the rapidity shift between the $pp$ CM and the CM of the string is
$\Delta = 1/2 \log (x_+/x_-)$. We then have

\begin{equation} \label{14e} N_1^{qq-q}(s,y) = \int_0^1 \int_0^1 dx_+ \
dx_- \ \rho_1^{qq}(x_+) \rho_1^{q} (x_-) {dN^{qq-q}\over dy} (y - \Delta
; s_{str}) \  .
\end{equation}

\noindent The subscript 1 in $N$ and $\rho$ indicates that there is
only one interaction (two strings). The momentum distribution functions
$\rho$ give the probability to find a quark or diquark in the proton
carrying a given momentum fraction. $dN^{qq-q}/dy$ is the rapidity
distribution of hadron $h$ in the CM of the $qq$-$q$ string obtained from
$q$ and $qq$ fragmentation functions~:

\begin{equation} \label{15e} {dN^{qq-q}(y - \Delta; s_{str}) \over dy} =
\left \{ \begin{array}{ll} \bar{x}_h D_{qq \to h} (x_h) &\quad y \geq \Delta
\ ,\\ \bar{x}_h D_{q\to h}(x_h) &\quad y < \Delta \ , \end{array} \right .
\end{equation}

\noindent where

\begin{equation} \label{16e} x_h = \left | 2 \mu_h \sinh (y -
\Delta)/\sqrt{s_{str}}\right | \ , \quad \bar{x}_h = \left ( x_h^2 + 4
\mu_h^2/s_{str}\right )^{1/2} \ . \end{equation}

\noindent $\mu_h$ is the transverse mass of the detected particle $h$,
and $D_{q \to h}$ and $D_{qq\to h}$ are the quark and diquark
fragmentation functions.
Momentum distribution and fragmentation functions can be obtained from
Regge intercepts. Let us discuss first the former. In order to
determine the behaviour near $x = 0$ of the momentum distribution of a
quark in a proton, it is convenient to look at the diagram in Fig.~6.
As discussed in Section \ref{subsec5}, the square of this diagram (in
the sense of unitarity) gives a
contribution to the total cross-section $s^{\alpha_R(0)-1} = e^{\Delta
y (1 - \alpha_R(0))}$ where $\Delta y = y - y_{max}$. Here $y$ is the
quark rapidity and $y_{max}$ its maximal value. Recalling that $dy =
dx/x$, we obtain $\rho_p^q(x_q) \propto x_q^{-\alpha_R(0)} =
1/\sqrt{x_q}$ as $x_q \to 0$. In order to determine its behaviour as
$x_q \to 1$, we have to use the momentum conservation $x_q + x_{qq} =
1$ (see Fig.~9). Thus, in order to have $x_q \to 1$ it is necessary
that $x_{qq} \to 0$. The corresponding Regge exchange in the
$t$-channel consists of two quarks and two antiquarks. Such a state is
called a baryonium and the corresponding Regge intercept is known
experimentally to be $- 1.5 \pm 0.5$. Taking the product of $x \to 0$
and $x \to 1$ behaviours we obtain

\begin{equation} \label{17e}
\rho_1^q(x_q) = \rho_1^{qq}(x_{qq}) = C x_q^{-1/2} \ x_{qq}^{1.5} \
\delta \left (1 - x_q - x_{qq}\right ) = C {1 \over \sqrt{x_q}} (1 -
x_q)^{1.5}
\end{equation}

\noindent $C$ is a constant determined from the normalization to unity.
We see from Eq. (\ref{17e}) that, in average, the quark is slow and the
diquark fast. \par

In order to generalize Eq. (\ref{17e}) to the case of $n$ inelastic
interactions ($2n$ strings), we just take the product of factors giving
the $x \to 0$ behaviour of each constituent, times a $\delta$-function
of momentum conservation. The
momentum distribution function $\rho_n(x)$ of each individual
constituent is then obtained by integrating over the $x$-values of the
other $2n-1$ constituents\footnote{Taking the same $1/\sqrt{x}$
behaviour for both valence and
sea quarks \cite{9r} \cite{10r}, these integrals can be performed
analytically and one gets~: $ \rho_n^q(x) = C_n x^{-1/2} (1-
x)^{n+1/2}$ and $\rho_n^{qq}(x) = C'_n x^{1.5} (1 - x)^{n-3/2}$, with
$C_n = \Gamma (n+2) /\Gamma (1/2) \Gamma (n+3/2)$ and $C'_n = \Gamma
(n+2)/\Gamma (5/2) \Gamma(n-1/2)$.}. In this way the behaviour $x \to 0$
is unchanged, whereas the power of $1 - x$ increases with $n$, due to
momentum conservation. Indeed, the average momentum fraction taken by
each constituent decreases when the number of produced strings
increases. Obviously in the case of $n$ inelastic collisions Eq.
(\ref{14e}) is still valid with $\rho_1$ replaced by $\rho_n$. All
details can be found in \cite{7r} \cite{9r}.\par

The same Regge model considerations allow to determine the $x_h \to 1$
behaviour of the fragmentation functions. Writing

\begin{equation} \label{18e}
\bar{x}_h \ D_{i\to h}(x_h) \propto \left ( 1 - x_h\right )^{\beta_i^h}
\end{equation}

\noindent for the fragmentation function of constituent $i$ into hadron
$h$ (see Eq. (\ref{15e})), one finds \cite{22r} $\beta_i^h = -
\alpha_{k\bar{k}}(0) + \lambda$ where $\alpha_{k\bar{k}}(0)$ is the
intercept of the $(k\bar{k})$ Regge trajectory, $k$ is the system of
leftover (spectator) constituents and $\lambda$ is a constant resulting
from transverse momentum integrations and estimated to be $\lambda \sim {1
\over 2}$ \cite{22r}. For example, for the fragmentation $u \to \pi^+$,
the system $k$ is a $d$-quark and $\beta_u^{\pi^+} = -
\alpha_{d\bar{d}}(0) + \lambda$. Likewise, for the fragmentation of a
$ud$ diquark into a proton we have $\beta_{ud}^p = -
\alpha_{u\bar{u}}(0) + \lambda$ and for that of a $ud$ diquark into
$\Lambda$, $\beta_{ud}^{\Lambda} = - \alpha_{s\bar{s}} (0) + \lambda$.
Here $\alpha_{u\bar{u}}(0) = \alpha_{d\bar{d}}(0) = \alpha_R(0) = 1/2$ and
$\alpha_{s\bar{s}}(0) = \alpha_{\phi} (0) = 0$. This gives a different
behaviour of the $p$ and $\Lambda$ inclusive spectrum which is observed
experimentally. \par

Note that in writing Eq. (\ref{13e}) we have assumed that individual
strings are independent from each other. In this way, the hadronic
spectra of a given graph are obtained by adding up the corresponding
ones for the individual strings. This leads to a picture, in which, for
any individual graph, particles are produced with only short-range (in
rapidity) correlations. Long-range correlations (and a broadening of
the multiplicity distributions) are due to fluctuations in the number
of strings, i.e. to the superposition of different graphs with their
corresponding weights. This gives a simple and successful description
of the data in hadron-hadron and hadron-nucleus interactions [7-10].

\subsection{Nucleus-Nucleus Interactions}
\label{subsec3.2}
The generalization of Eq. (\ref{3e}) to nucleus-nucleus collisions is
rather straighforward. For simplicity let us consider the case of $AA$
collisions and let $n_A$ and $n$ be the average number of participants
of each nucleus and the average number of binary $NN$ collisions,
respectively\footnote{$n_A(b) = \int d^2s A T_A(s) \left [ 1 - \exp
(-\sigma_{pp} A T_A (b-s)\right ]/\sigma_{AA}(b)$ and $n(b) =
\sigma_{pp} \int d^2s A^2 T_A(s) T_A(b-s) ]/\sigma_{AA}(b) =
\sigma_{pp}A^2T_{AA}(b)/\sigma_{AA}(b)$. These expressions can be
obtained in the Glauber model as follows. One has to generalize Eq.
(\ref{8e}) to the case of $AB$ collisions. The corresponding
cross-sections $\sigma_{n_A,n_B,n}(b)$ depend on three indices~:
$n_A(n_B)$ is the number of participants of nucleus $A(B)$ and $n$ is
the number of $NN$ collisions. Then, $n_A(b) =
\sum\limits_{n_A, n_B,n} n_A \sigma_{n_A,n_B,n}(b)/\sum\limits_{n_A,n_B,n}
\sigma_{n_A,n_B,n}(b)$ and $n(b) = \sum\limits_{n_A,n_B,n} n
\sigma_{n_A,n_B,n}(b)/\sum\limits_{n_A,n_B,n}
\sigma_{n_A,n_B,n}(b)$.}. At fixed impact parameter $b$, we have
\cite{23r} \cite{7r}
\begin{eqnarray}
\label{19e}
&&{dN^{AA} \over dy}(b) = n_A(b) \left [ N_{\mu (b)}^{qq-q_v}(y) +
N_{\mu (b)}^{q_v-qq}(y) + (2 k - 2) N_{\mu
(b)}^{q_s-\overline{q}_s}(y) \right ] \nonumber \\
&&+ (n(b) - n_A(b)) \ 2 k \ N_{\mu (b)}^{q_s-\overline{q}_s}(y) \ .
\end{eqnarray}

\noindent The physical meaning of Eq. (\ref{19e}) is quite obvious.
The expression in brackets corresponds to a $NN$ collision. Since $n_A$
nucleons of each nucleus participate in the collision, this expression
has to be multiplied by $n_A$. Note that in Eq. (\ref{13e}) the average
number of collisions is $k$ -- and the number of strings $2k$. In the
present case the total average number of collisions is $kn$ -- and the
number of strings $2kn$. The second term in Eq. (\ref{2e}) is precisely
needed in order to have the total number of strings required by the
model. Note that there are $2n_A$ strings involving the valence quarks
and diquarks of the participating nucleons. The remaining strings are
necessarily stretched between sea quarks and antiquarks. The value of
$\mu (b)$ is given by $\mu (b) = k \nu (b)$ with $\nu (b) =
n(b)/n_A(b)$, $\mu (b)$ represents the total average number of
inelastic collisions suffered by each nucleon. Actually, Eq.
(\ref{19e}) is an approximate expression, involving the same
approximation as in Eq. (\ref{13e}). The exact expression can be found
in \cite{23r} \cite{7r}. \par

We see from Eq. (\ref{19e}) that $dN^{AA}/dy$ is obtained as a linear
combination of the average number of participants and of binary
collisions. The coefficients are determined within the model and depend
on the impact parameter via $\mu (b)$. As discussed in Section
\ref{subsec3.1} the average invariant mass of a string containing a
diquark at one end is larger than the one of a $q$-$\overline{q}$
string since the average momentum fraction taken by a diquark is larger
than that of quark. It turns out that the same is true for the central
plateau, i.e.~: $N^{qq-q}(y^* \sim 0) > N^{q-\overline{q}}(y^* \sim
0)$. Let us now consider two limiting cases~:

\begin{equation}
\label{20e}
{\rm If} \ N^{q_s-\overline{q}_s}(y^* \sim 0) \ll N^{qq-q_v} (y^* \sim 0)\ , \
{\rm then}\ {dN^{AA} \over dy}(y^* \sim 0) \sim n_A \sim A^1 \end{equation}

\begin{equation}
\label{21e}
{\rm If} \ N^{q_s-\overline{q}_s}(y^* \sim 0) \sim N^{qq-q_v} (y^* \sim 0)\ , \
{\rm then}\ {dN^{AA} \over dy}(y^* \sim 0) \sim n \sim A^{4/3} \ .
\end{equation}

\noindent In the first case we obtain a proportionality in the number
of participants $n_A$ whereas in the second case we obtain a
proportionality in the number of binary collisions. Since $dN^{AA}/dy
\equiv (1/\sigma_{AA}) d\sigma^{AA}/dy$, the latter result implies that
$d\sigma^{AA}/dy \sim A^2$, i.e. all unitarity corrections cancel and
we obtain the same result as in the impulse approximation (Born term
only). This result is the AGK cancellation discussed in Section
\ref{sec:2}. It implies that, for the inclusive cross-section, soft and
hard processes have the same $A$-dependence. However, as discussed in
Section \ref{sec:2} the AGK cancellation is violated by diagrams
related to the diffraction production of large-mass states. These
diagrams give rise to shadowing corrections. Their effect is very
important in nuclear collisions since they are enhanced by $A^{1/3}$
factors.

\subsection{Charged Particle Multiplicities}
\label{subsec3.3}
At SPS energies the limit given by Eq. (\ref{21e}) is not reached, and
Eq. (\ref{19e}) leads to an $A$ dependence of $dN^{AA}/dy$ at $y^* \sim
0$ in $A^{\alpha}$ with $\alpha$ only slightly above unity. ($\alpha
\sim 1.08$ between 2 and 370 participants). On the other hand,
shadowing corrections are small due to phase space limitations. The
results \cite{21r} for $Pb$ $Pb$ collisions at $\sqrt{s} = 17.3$~GeV
are shown in Fig.~10. We see that both the absolute values and the
centrality dependence are well reproduced. When the energy increases,
Eq. (\ref{21e}) shows that the value of $\alpha$ should increase
towards $4/3$, in the absence of shadowing corrections. However, the
effect of the latter is increasingly important and, as a result, the
value of $\alpha$ varies little with $s$. At $\sqrt{s} = 130$~GeV,
without shadowing corrections the $A$-dependence is $A^{\alpha}$, with
$\alpha \sim 1.27$ in the same range of $n_{part}$ -- a value which is
not far from the maximal one, $\alpha = 4/3$ from Eq. (\ref{21e}). With
the shadowing corrections taken into account, the $A$-dependence is
much weaker (lower line
of the shaded area in Fig.~11) \cite{21r}. The $A$-dependence is now
$A^{\alpha}$ with $\alpha \sim 1.13$ -- always in the range of
$n_{part}$ from 2 to 370. As we see, the increase of $\alpha$ from SPS
to RHIC energies is rather small. This value of $\alpha$ is predicted
to change very little between RHIC and LHC, where $\alpha \approx 1.1$.
For, the increase from $\alpha \sim 1.27$ to $\alpha \sim 4/3$ obtained
in the absence of shadowing is compensated by an increase in the
strength of the shadowing corrections, leaving the effective value of
$\alpha$ practically unchanged. This implies that $dN/dy$ at $y^*
\sim 0$ in central $Au$ $Au$
collisions will increase by a factor $2 \div 2.5$ between RHIC and
LHC. This increase is slightly smaller than the corresponding increase
of $d\sigma/dy$ in $pp$ collisions.

\section{Nuclear Stopping}
\label{sec:4}
In $pp$ collisions the net proton ($p$-$\overline{p}$) distribution is
large in the fragmentation regions and has a deep minimum at
mid-rapidities. In contrast to this situation a much flatter
distribution has been observed \cite{24r} in central $Pb$ $Pb$
collisions at CERN-SPS. In view of that, several authors have
claimed that the stopping in heavy ion collisions is anomalous, in the
sense that it cannot be reproduced with the same mechanism (and the
same values of the parameters) used to describe the $pp$ data. In a
recent paper \cite{25r} it has been shown that this claim is not correct.\par

In the model described in the previous section, the net baryon can be
produced directly from the fragmentation of the diquark. Another
possibility is that the diquark splits producing a leading meson in the
first string break-up and the net baryon is produced in a further
break-up. Clearly, in the first case, the net baryon distribution will
be more concentrated in the fragmentation region than in the second
case. The corresponding rapidity distributions are related to the
intercepts of the relevant Regge trajectories, $\alpha_{qq}$ and
$\alpha_q$, respectively, i.e. they are given by $e^{\Delta y(1 -
\alpha )}$ (see Section \ref{subsec3.1}). Here
$\Delta y$ is the difference between the rapidity of the produced net
baryon and the maximal one. In the case of the first component, in
order to slow down the net baryon it is necessary to slow down a
diquark. The corresponding Regge trajectory is called baryonium and its
intercept is known experimentally to be $\alpha_{qq} \equiv
\alpha_{qq\bar{q}\bar{q}}(0) = - 1.5 \pm 0.5$ (see Section
\ref{subsec3.1}).
For the second component, where a valence quark is slowed down, we take
$\alpha_q \equiv \alpha_{q\bar{q}}(0) = \alpha_R(0) = 1/2$
\footnote{There is a third possibility in which the
net-baryon transfer in rapidity takes place without valence quarks
(string junction or gluonic mechanism) with intercept either
$\alpha_{SJ} = 1/2$ \cite{26r} or $\alpha_{SJ} = 1$ \cite{27r}. We find
no evidence for such a component from the existing $pp$ and $AA$ data.
Its smallness could be related to the fact that it produces an extra
string of hadrons and, thus, does not correspond to the dominant
topology in the large $N$ expansion.}. \par

In this way we arrive to the following two component model for net
baryon production $B_i - \overline{B}_i$ (where $i$ denotes the
baryon species) out
of a single nucleon

\begin{eqnarray}
\label{22e}
&&{dN^{B_i-\bar{B}_i} \over dy}(y,b) = I_2^i a\ C_{\mu(b)}\
Z_+^{1-\alpha_{q}(0)}(1 - Z_+)^{\mu
(b) - 3/2 + n_{sq}(\alpha_{R}(0)-\alpha_{\phi}(0))} \nonumber \\
&&+ I_1^i  (1 - a) C'_{\mu(b)} \ Z_+^{1- \alpha_{qq} (0)} \times (1 -
Z_+)^{\mu (b) - 3/2 + c + n_{sq}(\alpha_{R}(0) - \alpha_{\phi}(0))}
\end{eqnarray}

\noindent where $n_{sq}$ is the number of strange quarks in the
hyperon, $\alpha_{R}(0) = 1/2$, $\alpha_{\phi}(0) = 0$, $Z_+ =
(e^{y-y_{max}})$, $y_{max}$ is the maximal value of the baryon rapidity
and $\mu (b)$ is the average number of inelastic collisions suffered by
the nucleon at fixed impact parameter $b$ (see Section
\ref{subsec3.2}). The constants $C_{\mu}$ and $C'_{\mu}$ are
normalization constants required by baryon number
conservation\footnote{$C_{\nu} = \Gamma (a+b)/\Gamma (a) \Gamma (b)$
with $a = 1 - \alpha_q(0)$ and $b = \mu (b) - 1/2 + n_{sq}(\alpha_R(0)
- \alpha_{\phi}(0))$~; $C'_{\nu} = \Gamma (a' + b')/ \Gamma (a') \Gamma
(b')$ with $a' = 1 - \alpha_{qq}(0)$ and $b' = \mu (b) - 1/2 + c +
n_{sq}(\alpha_R(0) - \alpha_{\phi} (0)$).}. The small $Z$ behaviour is
controlled by the corresponding intercept. The factor $(1 - Z_+)^{\mu
(b) - 3/2}$ gives the $Z \to 1$ behaviour of the diquark momentum
distribution function in the case of $\mu$ inelastic collisions (see
footnote 5). Following conventional Regge rules an extra
$\alpha_{R}(0)- \alpha_{\phi}(0) = 1/2$ is added to the power of $1 -
Z_+$ for each strange quark in the hyperon (see Section
\ref{subsec3.1}, Eq. (\ref{18e})). \par

The weights $I_2^i$ and $I_1^i$ allow to determine the relative yields
of the different baryon and antibaryon species. They are computed in
Appendix A using simple quark counting rules. \par

The fraction, $a$, of the first component is treated as a free
parameter. The same for the parameter $c$ in the second component -- which
has to be determined from the shape of the (non-diffractive) proton
inclusive cross-section in the baryon fragmentation region. It can be
seen from Eq. (\ref{22e}) that stopping increases with $\mu (b)$, i.e.
with the total number of inelastic collisions suffered by each nucleon.
This effect is present in the two terms of (\ref{22e}) and is a
consequence of energy conservation. The question is whether this
``normal'' stopping is sufficient to reproduce the data. In other words
whether the data can be described with a universal value of $a$,
i.e. independent of $\mu$ and the same for all reactions.  \par

The formulae to compute net baryon production in $pp$ or $AA$
collisions can be obtained from Eq. (\ref{22e}) in a straightforward
way. Thus, in $AA$ collisions we have~: $dN^{AA\to B_i-\bar{B}_i}/dy
(y^*, b) = n_A(b) [dN^{B_i-\bar{B}_i}/dy (y^*, b) +
dN^{B_i-\bar{B}_i}/dy (-y^*,b)]$. \par

Note that in the formalism above, baryon quantum number is exactly
conserved. Note also that shadowing corrections are not present.
Indeed, as explained in Section \ref{sec:2}, these corrections affect
only the term proportional to the number of binary collisions, which is
not present for net baryon production.

A good description of the data on the rapidity distribution of $pp
\to p - \overline{p}
+ X$ both at $\sqrt{s} = 17.2$~GeV and $\sqrt{s} = 27.4$~GeV is
obtained from Eq.
(\ref{22e}) with $a = 0.4$, $c = 1$, $\alpha_{q} = 1/2$ and $\alpha_{qq}
= -1$. The results are shown in Table \ref{tab:1} at three different
energies, and compared with
the data. As we see the agreement is reasonable. As it
is well known, a pronounced minimum is present at $y^* = 0$. There is also a
substantial decrease of the mid-rapidity yields with increasing
energy. Also, the
mid-rapidity distributions get flatter with increasing energy since
the net proton
peaks are shifted towards the fragmentation regions. \par

It is now possible to compute the corresponding net baryon production
in heavy ion collisions and to check whether the data can be described
with Eq. (\ref{22e}) using the same set of parameters as in $pp$. The
results for net protons $(p-\overline{p})$ in central $Pb$ $Pb$
collisions at $\sqrt{s} = 17.2$~GeV and central $Au$ $Au$ collisions at
$\sqrt{s} = 200$~GeV are shown in Fig.~14. We see a dramatic change in
the shape of the rapidity distribution between the two energies, which
is reasonably described by the model. Therefore, we conclude that there
is no need for a new mechanism in $AA$ collisions.

\section{Hyperon and Antihyperon Production}
\label{sec:5}
Strange particle production, in particular, of multistrange hyperons,
has been proposed as a signal of Quark Gluon Plasma formation. Flavor
equilibration is very efficient in a plasma due to large gluon
densities and low thresholds \cite{4r}. Moreover, the increase of the
relative yields of strange particles in central $AA$ collisions as
compared to $pp$ can be understood as a consequence of the necessity of
using the canonical ensemble in small size systems $(pp)$ -- rather
than the grand canonical one. The exact conservation of quantum numbers
in the former leads to a reduction of $s\bar{s}$ pair production, as
compared to the latter \cite{5r}. \par

An analysis of the results at SPS in the framework of the present
model has been presented in
\cite{28r}. In the following we concentrate on RHIC results. \par

A general result in DPM is that the ratios $B/h^-$ and
$\overline{B}/h^-$ of baryon and antibaryon yields over negatives
decrease with increasing centralities. This is easy to see from Eq.
(\ref{19e}). The production from $q_s$-$\overline{q}_s$ strings scales
with the number of binary collisions. These strings have a smaller
(average) invariant mass than the $qq$-$q$ strings and, thus, are more
affected by the thresholds needed for $B\overline{B}$ pair production.
As a consequence, the centrality dependence of $B$ and $\overline{B}$
production will be smaller than the one of negatives.  The effect is
rather small at RHIC energies. However, it
is sizable and increases with the mass of the produced baryon. In
contrast with this situation, the data for $\Lambda$'s show no such
decrease and an increase is present for $\Xi$ production. Data on
$\Omega$ production are not yet available. However, SPS data clearly
show a hierachy in the sense that the enhancement of baryon production
increase with the mass (or strange quark content) of the produced
baryon.

The only way out we have found is to give up the assumption of string
independence. Until
now we have assumed that particles produced in different strings are
independent from
each other. In the following we allow for some final state
interactions between comoving
hadrons or partons. We proceed as follows.\par

The hadronic densities obtained in Section 2 are used as initial
conditions in the gain
and loss differential equations which govern final state interactions. In the
conventional derivation \cite{29r} of these equations, one uses
cylindrical space-time
variables and assumes boost invariance. Furthermore, one assumes that
the dilution in
time of the densities is only due to longitudinal motion, which leads
to a $\tau^{-1}$
dependence on the longitudinal proper time $\tau$. These equations
can be written
\cite{29r} \cite{28r}

\begin{equation}
\label{23e}
\tau \ {d\rho_i \over d \tau} = \sum_{k\ell} \sigma_{k\ell} \ \rho_k
\ \rho_{\ell} -
\sum_k \sigma_{ik}\ \rho_i \ \rho_k \ . \end{equation}

\noindent The first term in the r.h.s. of (\ref{23e}) describes the
production (gain) of particles of type $i$ resulting from the
interaction of particles $k$ and $\ell$. The second term describes the
loss of particles of type $i$ due to its interactions with particles of
type $k$. In Eq. (\ref{23e}) $\rho_i = dN_i/dyd^2s(y,b)$ are the
particles yields per unit rapidity and per unit of transverse area, at
fixed impact parameter. They can be obtained from the rapidity
densities, Eq. (\ref{19e}), using the geometry, i.e. the $s$-dependence
of $n_A$ and $n$. The procedure is explained in detail in \cite{30r}
where the pion fragmentation functions are also given. Those of kaons
and baryons can be found in \cite{31r}. These fragmentation functions
are obtained using the procedure sketched at the end of Section
\ref{subsec3.1} (see Eq. (\ref{18e})). $\sigma_{k\ell}$ are the
corresponding cross-sections averaged over the momentum distribution of
the colliding particles. \par

Equations (\ref{23e}) have to be integrated from initial time $\tau_0$
to freeze-out time $\tau_f$. They are invariant under the change $\tau
\to c \tau$ and, thus, the result depends only on the ratio
$\tau_f/\tau_0$. We use the inverse proportionality between proper time
and densities and put $\tau_f/\tau_0 = (dN/dyd^2s(b))/\rho_f$. Here the
numerator is given by the DPM particles densities. We take $\rho_f =
[3/\pi R_p^2](dN^-/dy)_{y^* \sim 0} = 2$~fm$^{-2}$, which corresponds
to the density of charged and neutrals per unit rapidity in a $pp$
collision at $\sqrt{s} = 130$~GeV. This density is about 70 \% larger
than at SPS energies. Since the corresponding increase in the $AA$
density is comparable, the average duration time of the interaction
will be approximately the same at CERN SPS and RHIC -- about 5 to 7
fm.\par

Next, we specify the channels that have been taken into account in
our calculations.
They are

\begin{equation}
\label{24e}
\pi N \stackrel{\rightarrow}{\leftarrow} K \Lambda (\Sigma)\ , \quad
\pi \Lambda (\Sigma )
\stackrel{\rightarrow}{\leftarrow} K \Xi \ , \quad \pi \Xi
\stackrel{\rightarrow}{\leftarrow} K \Omega  \ .\end{equation}

\noindent We have also taken into account the strangeness exchange reactions

\begin{equation}
\label{25e}
\pi \Lambda (\Sigma ) \stackrel{\rightarrow}{\leftarrow} K N\ , \quad \pi \Xi
\stackrel{\rightarrow}{\leftarrow} K \Lambda (\Sigma ) \ , \quad \pi
\Omega  \stackrel{\rightarrow}{\leftarrow}
K \Xi \end{equation}

\noindent as well as the channels corresponding to (\ref{24e}) and
(\ref{25e}) for
antiparticles. We have taken
$\sigma_{ik} = \sigma = 0.2$~mb, i.e. a single value for all
reactions in (\ref{24e}) and
(\ref{25e}) -- the same value used in ref. \cite{28r} to describe the
CERN SPS data.
\par

Before discussing the numerical results and the comparison with
experiment let us
examine the qualitative effects of comovers interaction. As explained
in the beginning of
this Section, without final state interactions all ratios $K/h^-$, $B/h^-$
and $\overline{B}/h^-$ decrease with increasing centrality. The final
state interactions (\ref{24e}), (\ref{25e}) lead to a gain of strange
particle
yields.  The reason for this is the following. In the first direct
reaction (\ref{24e}) we
have $\rho_{\pi} > \rho_K$, $\rho_N > \rho_{\Lambda}$, $\rho_{\pi}
\rho_N \gg \rho_K
\rho_{\Lambda}$. The same is true for all direct reaction
(\ref{24e}). In view of that,
the effect of the inverse reactions (\ref{24e}) is small. On the
contrary, in all
reactions (\ref{25e}), the product of densities in the initial and
final state are
comparable and the direct and inverse reactions tend to compensate
with each other.
Baryons with the largest strange quark content, which find themselves
at the end of the
chain of direct reactions (\ref{24e}) and have the smallest yield
before final state
interaction, have the largest enhancement. Moreover, the gain in the
yield of strange
baryons is larger than the one of antibaryons since $\rho_B >
\rho_{\overline{B}}$.
Furthermore, the enhancement of all baryon species increases with
centrality, since the
gain, resulting from the first term in Eq. (\ref{23e}), contains a
product of densities
and thus, increases quadratically with increasing centrality.\par

In Fig.~15a-15d we show the rapidity densities of $B$, $\overline{B}$
and $B - \overline{B}$ versus $h^- = dN^-/d\eta = (1/1.17) dN^-/dy$
\cite{31r} and
compare them with available data [32-34]. We would like to stress that
the results for $\Xi$ and $\overline{\Xi}$ were given \cite{31r} before
the data \cite{34r}. This is an important success of our approach.\par

In first approximation, the yields of $p$, $\overline{p}$, $\Lambda$
and $\overline{\Lambda}$ yields over $h^-$ are independent of
centrality. Quantitatively, there is a slight decrease with centrality
of $p/h^-$ and $\overline{p}/h^-$ ratios, a slight increase of $\Lambda
/h^-$ and $\overline{\Lambda}/h^-$ and a much larger increase for $\Xi$
($\overline{\Xi})/h^-$ and $\Omega$ ($\overline{\Omega})/h^-$. This is
better seen in Figs.~16a and 16b where we plot the yields of $B$ and
$\overline{B}$ per participant normalized to the same ratio for
peripheral collisions versus $n_{part}$. The enhancement of $B$ and
$\overline{B}$ increases with the number of strange quarks in the
baryon. This increase is comparable to the one found at SPS between
$pA$ and central $Pb$ $Pb$ collisions. (In the statistical approach
\cite{5r}, the enhancement of $B$ and $\overline{B}$ relative to $pp$
decreases with increasing energy. This may allow to distinguish between
the two approaches). \par

The ratio $K^-/\pi^-$ increases by 30~\% in the same centrality range,
between 0.11 and 0.14 in agreement with present data. The ratios
$\overline{B}/B$ have a mild decrease with centrality of about 15~\%
for all baryon species -- which is also seen in the data. Our values
for $N^{ch}/N_{max}^{ch} = 1/2$ are~: $\overline{p}/p = 0.69$,
$\overline{\Lambda}/\Lambda = 0.74$, $\overline{\Xi}/\Xi = 0.79$,
$\Omega/\overline{\Omega} = .83$, to be compared with the measured
values \cite{35r}~:

$$\overline{p}/p = 0.63 \pm 0.02 \pm 0.06 \quad , \quad
\overline{\Lambda}/\Lambda = 0.73 \pm 0.03 \quad , \quad
\overline{\Xi}/\Xi = 0.83 \pm 0.03 \pm 0.05 \ .$$

\noindent The ratio $K^+/K^- = 1.1$ and has a mild increase with
centrality, a feature also seen in the data. \par

Note that a single parameter has been adjusted in order to determine
the absolute yields of $B\overline{B}$ pair production, namely the
$\overline{p}$ one -- which has been adjusted to the experimental
$\overline{p}$ value for peripheral collisions. The yields of all other
$B\overline{B}$ pairs has been determined using the quark counting
rules given in Appendix A. \par

Although the inverse slopes (``temperature'') have not been discussed
here, let us note that in DPM they are approximately the same for all
baryons and antibaryons both before and after final state interaction
-- the effect of final state interaction on these slopes being rather
small \cite{36r}.

\section{$J/\psi$ Suppression} \label{sec:6} A most interesting
signature of the production of QGP is the suppression of resonance
production \cite{3r}. As a consequence of deconfinement, the resonances
are ``melted'', i.e. the bound state cannot be formed. More precisely,
as in the case of an ordinary plasma, the potential $V_0(r)$ is
screened (Debye screening), i.e. it is changed into $V_0(r) \exp
(-r/r_D(T))$. Here $r_D(T)$ is the Debye radius that decreases with
increasing temperature. When $r_D(T)$ becomes smaller than the hadronic
radius, the bound state cannot be formed. This idea is particulary
interesting in the case of the $J/\psi$ (a resonance consisting of a
charm quark and its antiquark). Indeed, it has been shown that the
melting of the $J/\psi$ occurs at temperature only slightly higher than
the critical temperature at which QGP is formed. Moreover, the
production of $c$-$\bar{c}$ pairs is very rare. If they cannot bind
together the $\bar{c}(c)$ will combine with a light quark (antiquark)
giving rise to a $\bar{D}(D)$ meson (open charm)\footnote{In what
follows we disregard the possibility of $c$ and $\bar{c}$ recombination
into a $J/\psi$.}.\par

The NA38-NA50 collaborations have observed a decrease of the ratio of
$J/\psi$ to dimuon (DY) cross-sections with increasing centrality in
$SU$ and $Pb$ $Pb$ collisions \cite{37r}. The same phenomenon has
been observed in
$pA$ collisions with increasing values of $A$. In this case, it is
interpreted as due to the interaction of the pre-resonant
$c\overline{c}$ pair with the nucleons of the nucleus it meets in its
path (nuclear absorption). [Indeed, the formation time of the
$J/\psi$ is longer and it is produced outside the nucleus.] As a
result of this interaction, the
$c\overline{c}$ pair is modified in such a way that, after interaction,
it has no projection into $J/\psi$ (a $D\overline{D}$ pair is produced
instead).  The corresponding cross-section is denoted $\sigma_{abs}$
(absorptive
cross-section). \par

The survival probability, $S_{abs}(b)$, of the $J/\psi$ in $pA$ collisions
can be easily calculated in the probabilistic Glauber model. One has

\begin{eqnarray} \label{26e} &&S_{abs}^A(b) =
\int_{-\infty}^{+\infty} dZ \rho_A(b,Z) \left . \left ( 1 - \sigma_{abs}
\int_Z^{+\infty} \rho_A (b, Z_1) dZ_1\right )^{A-1} \right / \nonumber \\
&&\int_{-\infty}^{+\infty} dZ \rho_A(b, Z) = {1
\over \sigma_{abs}AT_A(b)} \left \{ 1 - \left [ 1 - \sigma_{abs}
T_A(b) \right ] \right \}^A \ . \end{eqnarray}

\noindent Indeed, the $c\bar{c}$ pair is produced at a point of
coordinates $(b,Z)$ inside the nucleus, with probability proportional
to $\rho_A(b,Z)$. The term inside the parenthesis gives the probability
of non-absorption during its subsequent propagation through the
nucleus. Note that $S_{abs} = 1$ for $\sigma_{abs} = 0$. The
generalization of (\ref{26e}) to the case of
nucleus-nucleus interactions is rather straightforward. We have \cite{30r}

\begin{equation}
\label{27e}
S_{abs}^{AB} (b, s) = S_{abs}^A(s) \ S_{abs}^B (b - s) \ .
\end{equation}

The NA50 collaboration has shown that the $J/\psi$ suppression in $Pb$
$Pb$ collisions has an anomalous component, i.e. it cannot be
reproduced using nuclear absorption alone \cite{37r}. Two main
interpretations have
been proposed~: deconfinement and comovers interaction. The latter
mechanism has been described in Section \ref{sec:5} for hyperon
production. In the case of $J/\psi$ suppression, a single channel is
important namely $c\overline{c}$ (or $J/\psi$) interacting with comoving
hadrons and producing a $D\overline{D}$ pair. In this case, Eq.
(\ref{23e}) can be solved analytically. One obtains for the
expression of the survival
probability $S_{co}$ \cite{30r}

\begin{equation}
\label{28e}
S_{co}^{AB}(b,s) = \exp \left [ - \sigma_{co} {3 \over 2}
N_{yDT}^{co}(b,s) \ell n {\tau_f \over \tau_0} \right ] \ .
\end{equation}

\noindent Here $N_{yDT}^{co}(b, s)$ is the density of charged
particles in the rapidity region of the dimuon trigger (DT) $(0 < y^* <
1)$ computed from Eq. (\ref{19e}). The factor 3/2 takes care of the
neutrals. For $\tau_f/\tau_0$ in Eq. (\ref{28e}) we use the expression
given in Section \ref{sec:5}, with $\rho_f = 1.15$~fm$^{-2}$ at SPS
energies. Note that $S_{co}$ depends on a single parameter
$\sigma_{co}$, the effective cross-section for comovers
interaction.\par

The results \cite{38r} of the comovers interaction model are presented
in Fig.~17. The agreement with the data \cite{37r} is quite
satisfactory. There is a
single free parameter $\sigma_{co} = 0.65$~mb. The value of
$\sigma_{abs} = 4.5$~mb is determined \cite{38r} from the $pA$ data and
the absolute normalization (47) from the SU ones. \par

Predictions of the comovers
model \cite{38r} at RHIC energies are given in Figure 18. \par

In a deconfining approach one proceeds as follows \cite{39r}. One
assumes that the energy density of the produced system is
proportional to the density of participants $n_A(b,s)$ of nucleus $A$
in a $AA$ interaction. If $n_A (b,s) <
n_{crit}$ the $J/\psi$ is suppressed only due to ordinary nuclear
absorption with cross-section $\sigma_{abs}$. On the contrary, if
$n_A(b, s) \ \gsim \ n_{crit}$, the nuclear absorption formula is used
with $\sigma_{abs}$ infinity. In this way, no $J/\psi$ can survive
above the critical density. The deconfining approach leads to a
satisfactory description of $J/\psi$ suppression, with $n_{crit}$
treated as a free parameter\footnote{Both in the comovers model and
in the deconfining one, the
description of the $J/\psi$ suppression at very large transverse
energy, $E_T$, requires the introduction of $E_T$-fluctuations
\cite{30r} \cite{39r}.}. However, a quantitative analysis of the most
recent NA50 data \cite{37r} is still missing. On the other hand, the
centrality dependence of the average $p_T$ of $J/\psi$ is better
described in the comovers approach than in a deconfining scenario
\cite{40r}.

\section{Conclusions}
\label{sec:7}
In these lectures a description of microscopic string models of
hadronic and nucleus interactions has been presented. Consequences of
the model for charged particles multiplicities, net baryon production
(stopping), hyperon and antihyperon production (strangeness
enhancement) and $J/\psi$ suppression have been examined. \par

As a starting point we have assumed that particles produced in
different strings are independent, i.e. there is no ``cross-talk''
between strings. While this assumption works quite
well in $pp$ and $pA$ interactions, the $AA$ data can only be
described if final state interaction (comovers interaction) is
introduced. However, the corresponding cross-section turns out to be
rather small (a few tenths of a milibarn). Due to this smallness and to
the short duration time of final state interaction ($5\div 7$~fm) it is
unlikely that thermal equilibrium can be reached.\par

Of course it is not possible to reach a definite conclusion on this
important point. However, particle abundances not only do not allow to
conclude that equilibrium has been reached, but, on the contrary, their
centrality dependence tends to indicate that this is not the case. Let
us consider for instance $p$ and $\overline{p}$ production. In our
model, their yields are practically not affected by final state
interaction, i.e. they are practically the same assuming string
independence. Yet, the model reproduces the data, from very peripheral
to very central interaction. This success would be difficult to
understand in a QGP scenario in which for peripheral collisions (below
the critical density) there is strong, non-equilibrated,
$p\overline{p}$ annihilation, which becomes equilibrated for central
ones, above the critical density. More generally, the QGP scenario
would be strongly supported if some kind of threshold would be found in
the strange baryon yields around the critical energy density. At SPS
energies, evidence for such a threshold in the $\overline{\Xi}$ yield
was claimed by the NA57 collaboration based on preliminary data
\cite{41r}, but it is not seen in the more recent analysis \cite{42r}.
Moreover, the saturation at large centralities of the $B$ and
$\overline{B}$ yields per participants, shown by the WA97 data, has
also disappeared from the new data \cite{42r}, in agreement with the
predictions of the comovers interaction model \cite{38r}.
Unfortunately, these data only cover a limited range of centrality. In
contrast to this situation, the RHIC data explore the whole centrality
range from very peripheral to very central collisions and the
centrality dependence of the yields of $p$, $\Lambda$, $\Xi$ and their
antiparticles shows no structure whatsoever. Moreover, the yields of
$\Xi$ and $\overline{\Xi}$ per participant (as well as the ratios
$\Xi/h^-$ and $\overline{\Xi}/h^-$) do not seem to saturate at large
centralities. If the same happens for $\Omega$ and $\overline{\Omega}$
production (as predicted in our approach, Figs.~16) the case for QGP
formation from strange baryon enhancement will be quite weak.\par

Finally, it should be stressed that the final state interaction of
comovers in our approach is by no means a trivial hadronic effect.
Indeed, the interaction of comovers starts at the early times when
densities, as computed in DPM, are very large. In this situation the
comovers are not hadrons (there are several of them in the volume
normally occupied by one hadron, and, moreover, at these early times
hadrons are not yet formed). This is probably the reason why in our
approach the comover interaction cross-sections required to describe
the data are smaller than in hadron gas models where the final state
interaction starts only after hadron formation.

\begin{table}
\centering
\caption{Calculated values \cite{25r} of the rapidity distribution of
$pp \to p - \overline{p} + X$ at
$\sqrt{s} = 17.2$~GeV and 27.4 GeV ($k = 1.4$) and $\sqrt{s} =
130$~GeV ($k=2$). (In
order to convert $d\sigma/dy$ into $dN/dy$ a value of $\sigma =
30$~mb has been used).
For comparison with the nucleus-nucleus results, all values in this table have
been scaled by $n_A = 175$.}
\label{tab:1}
\begin{tabular}{cccc}
\hline\noalign{\smallskip}
$y^*$ &$pp \to p - \overline{p}$ &$pp \to p - \overline{p}$ &$pp \to
p - \overline{p}$ \\
&$\sqrt{s} = 17.2$ GeV &$\sqrt{s} = 27.4$ GeV &$\sqrt{s} = 130$ GeV \\
\hline\noalign{\smallskip}
0 &9.2 &6.5 &3.6 \\
& &$[6.3 \pm 0.9 ]$ & \\
1 &15.0 &9.3 &4.2 \\
&$[16.1 \pm 1.8]$ &$[9.6 \pm 0.9]$ & \\
1.5 &25.8 &14.6 &5.1 \\
&$[24.1 \pm 1.4]$ &$[15.4 \pm 0.9]$ & \\
2 &47.1 &26.2 &6.8 \\
&$[45.4 \pm 1.4]$ &$[27.7 \pm 0.9]$ & \\
\hline\noalign{\smallskip}
\end{tabular}
\end{table}

\newpage

\noindent {\large \bf Appendix A}
\label{A:app}
\vskip 1 truecm
In order to get the relative densities of each baryon and antibaryon
species we use
simple quark counting rules \cite{28r} \cite{31r}. Denoting the
strangeness suppression factor by
$S/L$ (with $2L+ S = 1$), baryons produced out of three sea quarks
(which is the case for pair
production) are given the relative weights

$$I_3 = 4L^3 : 4L^3 : 12L^2S : 3LS^2 : 3LS^2 : S^3 \eqno({\rm A}.1)$$

\noindent for $p$, $n$, $\Lambda + \Sigma$, $\Xi^0$, $\Xi^-$ and
$\Omega$, respectively. The
various coefficients of $I_3$ are obtained from the power expansion
of $(2L + S)^3$.\par

For net baryon production, we have seen in Section \ref{sec:4} that
the baryon can contain
either one or two sea quarks. The first case corresponds to direct
diquark fragmentation
described by the second term of Eq. (\ref{22e}). The second case
corresponds to diquark
splitting, described by the first term of (\ref{22e}). In these two
cases, the relative
densities of each baryon species are respectively given by

$$I_1 = L : L : S \eqno({\rm A}.2)$$

\noindent for $p$, $n$ and $\Lambda + \Sigma$, and

$$I_2 = 2L^2 : 2L^2 : 4LS : {1 \over 2} S^2 : {1 \over 2} S^2
\eqno({\rm A}.3)$$

\noindent for $p$, $n$, $\Lambda + \Sigma$, $\Xi^0$ and $\Xi^-$. The
various coefficients in
(A.2) and (A.3) are obtained from the power expansion of $(2L + S)$
and $(2L + S)^2$,
respectively.\par

In order to take into account the decay of $\Sigma^*(1385)$ into
$\Lambda \pi$, we
redefine the relative rate of $\Lambda$'s and $\Sigma$'s using the
empirical rule
$\Lambda = 0.6(\Sigma^+ + \Sigma^-$) -- keeping, of course, the total yield of
$\Lambda$'s plus $\Sigma$'s unchanged. In this way the normalization
constants of all
baryon species are determined from one of them. This constant,
together with the relative normalization of $K$ and $\pi$, are
determined from the data
for very peripheral collisions. In the calculations we use $S = 0.1$
$(S/L = 0.22)$.


\newpage

\centerline{\large\bf Figure captions}\par \vskip 5 truemm

\noindent{\bf Figure 1} : Phase diagram in statistical QCD \cite{1r}
\cite{2r}.\par \vskip 3 truemm

\noindent{\bf Figure 2} : Energy density versus temperature in lattice
QCD \cite{1r} at $\mu = 0$, for three light flavors (upper), two light
and one heavier (middle) and two light flavors (lower).\par \vskip 3
truemm

\noindent{\bf Figure 3} : Single (a), double (b), and multiple (c)
scattering diagrams in the eikonal model.\par \vskip 3 truemm

\noindent{\bf Figure 4} : The secondary Regge trajectories with highest
intercept.\par \vskip 3 truemm

\noindent{\bf Figure 5} : The mechanism of particle production in
$e^+e^-$ annihilation. The net of soft gluons and quark loops is only
shown here and in Fig. 10.\par \vskip 3 truemm

  \noindent{\bf Figure 6} : One string diagram in
$\overline{p}p$. \par \vskip 3 truemm

\noindent{\bf Figure 7} : Dominant two-chain (single
cut Pomeron) contributions to high energy $\pi^+$-proton collisions.
\par \vskip 3 truemm

\noindent{\bf Figure 8} : Dominant two-chain
contribution to proton-antiproton collisions at high energies (single
cut Pomeron).\par
\vskip 3 truemm

\noindent{\bf Figure 9} : Dominant two-chain diagram describing
multiparticle production in
high energy proton-proton collisions (single cut Pomeron). \par \vskip 3 truemm

\noindent{\bf Figure 10} : Single Pomeron exchange and its underlying
cylindrical topology.
This is the dominant contribution to proton-proton elastic scattering
at high energies.
\par \vskip 3 truemm

\noindent {\bf Figure 11} : Two cut Pomeron (four-chain) diagram for
proton-proton
collisions. \par \vskip 3 truemm

\noindent {\bf Figure 12} : The values of $dN^{ch}/dy$
per participant for $Pb$ $Pb$ collisions at $\sqrt{s} = 17.3$~GeV
computed \cite{21r} from Eq. (\ref{19e}), compared with WA98 data.
\par \vskip 3 truemm

\noindent {\bf Figure 13} : The values of $dN^{ch}/d\eta_{c.m.}/(0.5\
n_{part})$ for $Au$ $Au$ collisions at $\sqrt{s} = 130$~GeV computed
\cite{21r}
from Eq. (\ref{19e}) including shadowing corrections are given by the dark band
in between solid lines. The PHENIX data are also shown (black circles
and shaded area).\par \vskip 3 truemm

\noindent {\bf Figure 14} : Rapidity distribution of net protons
($p$-$\overline{p}$) for the 5~\% most central $Pb$ $Pb$ collisions at
SPS ($\sqrt{s} = 17.2$~GeV) and for the 10~\% most central $Au$ $Au$
ones at $\sqrt{s} = 200$~GeV, compared to data \cite{24r} \cite{43r}.
\par \vskip 3 truemm

\noindent {\bf Figure 15} : (a) Calculated values \cite{31r} of
$dN/dy$ of $p$ (solid
line) $\overline{p}$ (dashed line), and $p - \overline{p}$ (dotted
line) at mid rapidities, $|y^*| < 0.35$, are plotted as a function of
$dN_{h^-}/d\eta$, and compared with PHENIX data \cite{32r}~; (b) same for
$\Lambda$ and $\overline{\Lambda}$ compared with preliminary STAR data
\cite{33r}~; (c) same for $\Xi^-$ and $\overline{\Xi}^+$ compared to
preliminary STAR data \cite{34r}~; (d) same for $\Omega$ and
$\overline{\Omega}$. \par \vskip 3 truemm

\noindent {\bf Figure 16} : Calculated values \cite{31r} of the
ratios $B/n_{part}$ (a)
and $\overline{B}/n_{part}$ (b), normalized to the same ratio for
peripheral collisions ($n_{part} = 18$), plotted as a function of
$n_{part}$.\par \vskip 3 truemm

\noindent {\bf Figure 17} : The ratio of $J/\psi$ over $DY$ cross-sections
in $Pb$ $Pb$ collisions a 158 GeV/c versus $E_T$ obtained \cite{38r} in the
comovers interaction model with $\sigma_{abs} = 4.5$~mb and
$\sigma_{co} = 0.65$~mb. The absolute normalization is 47. The
preliminary data are from \cite{37r}.\par \vskip
3 truemm

\noindent {\bf Figure 18} : $dN/dy$ of $J/\psi$ (times branching ratio) in
$Au$ $Au$ collisions at $\sqrt{s} = 200$~GeV per nucleon and $y^* \sim
0$, scaled by the number of binary collisions $<nb>$, versus the number
of participants. The curves are obtained in the comovers model \cite{38r}
with $\sigma_{abs} = 0$, $\sigma_{co} = 0.65$~mb (upper) and
$\sigma_{abs} = 4.5$~mb, $\sigma_{co} = 0.65$~mb (lower). The absolute
normalization is arbitrary. An extra 20~\% suppression between $pp$ and
central $Au$ $Au$ is expected due to shadowing.

\newpage

\centerline{{\epsfxsize15cm \epsfbox{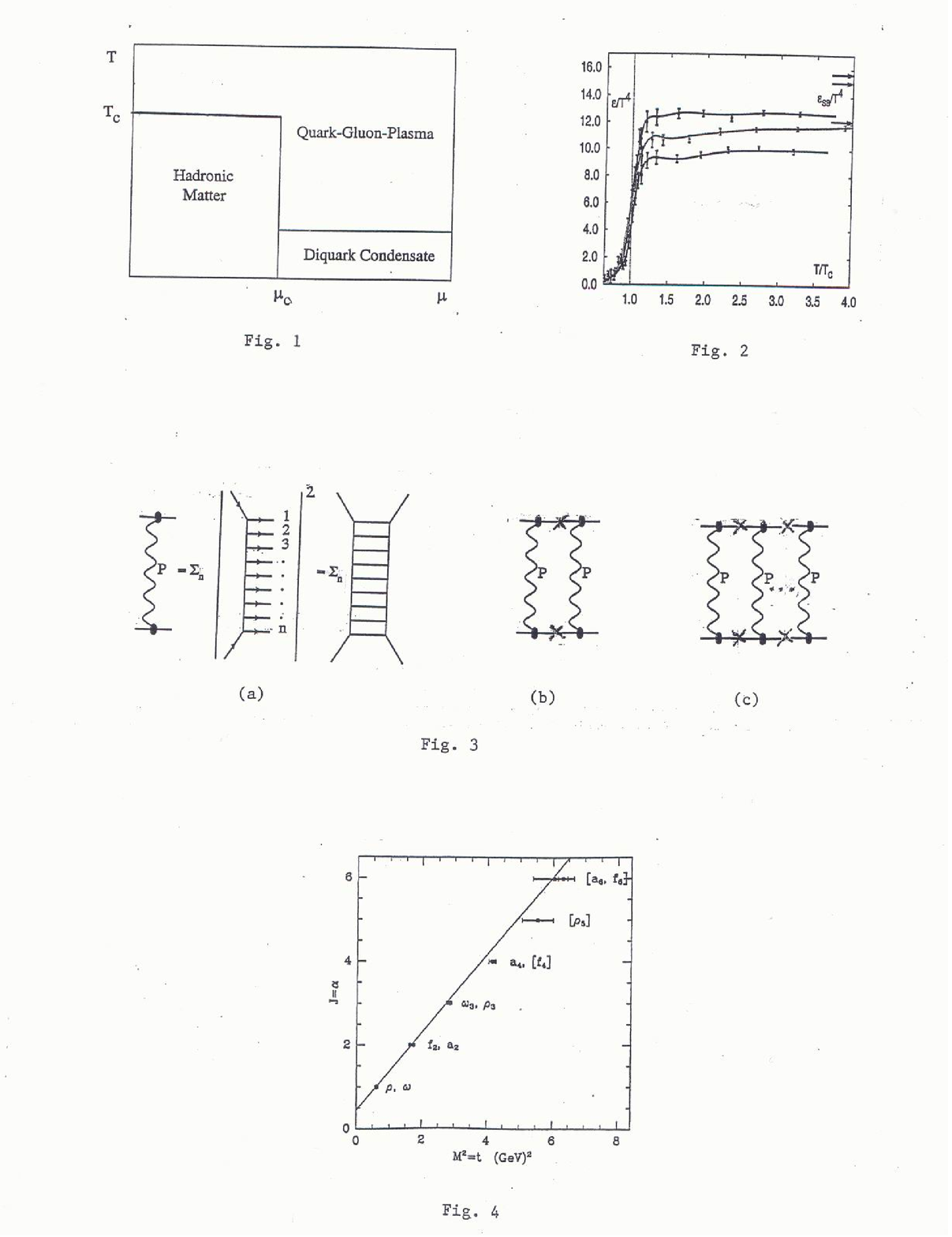}}}

\centerline{{\epsfxsize15cm \epsfbox{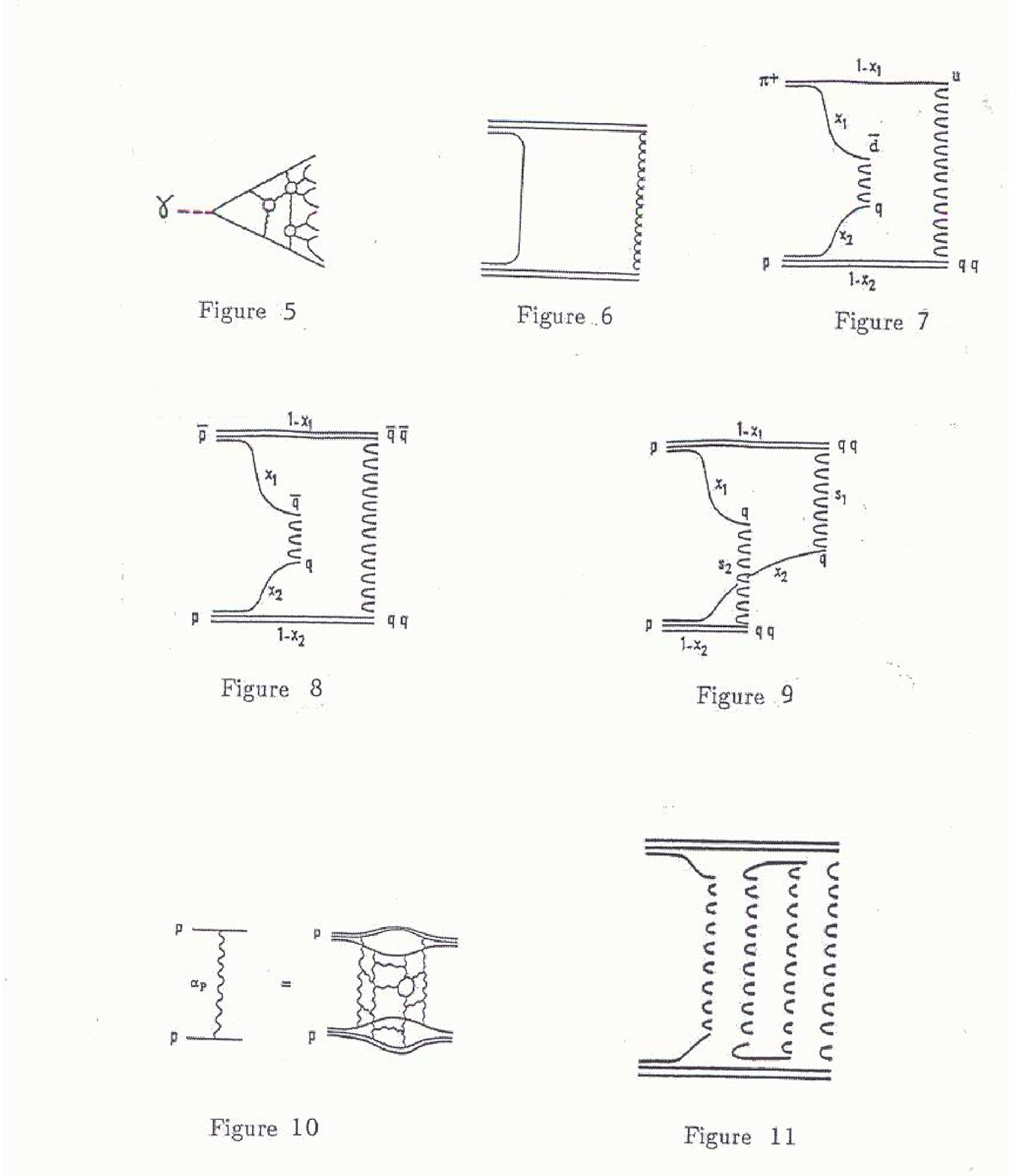}}}

\centerline{{\epsfxsize15cm \epsfbox{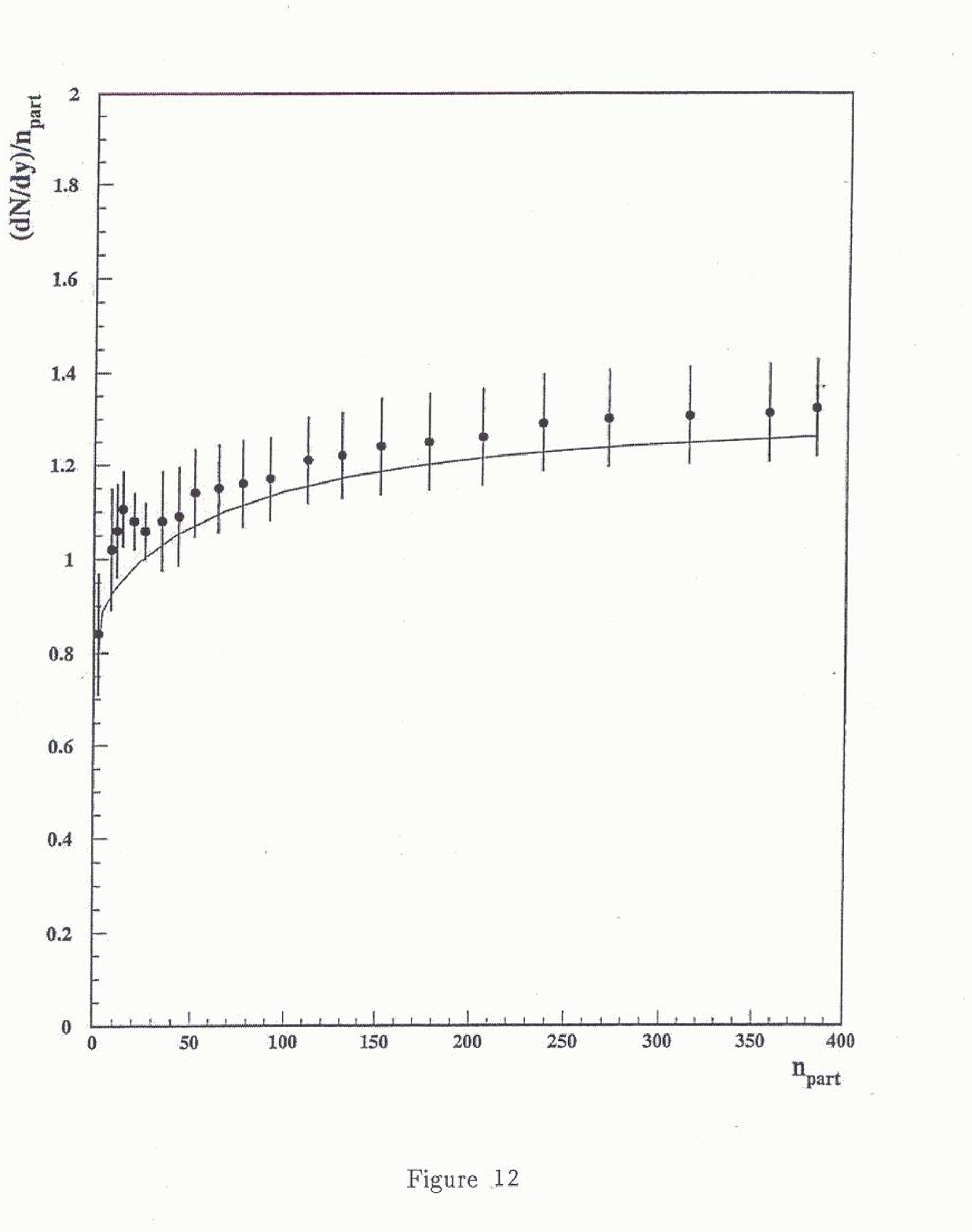}}}

\centerline{{\epsfysize21cm \epsfbox{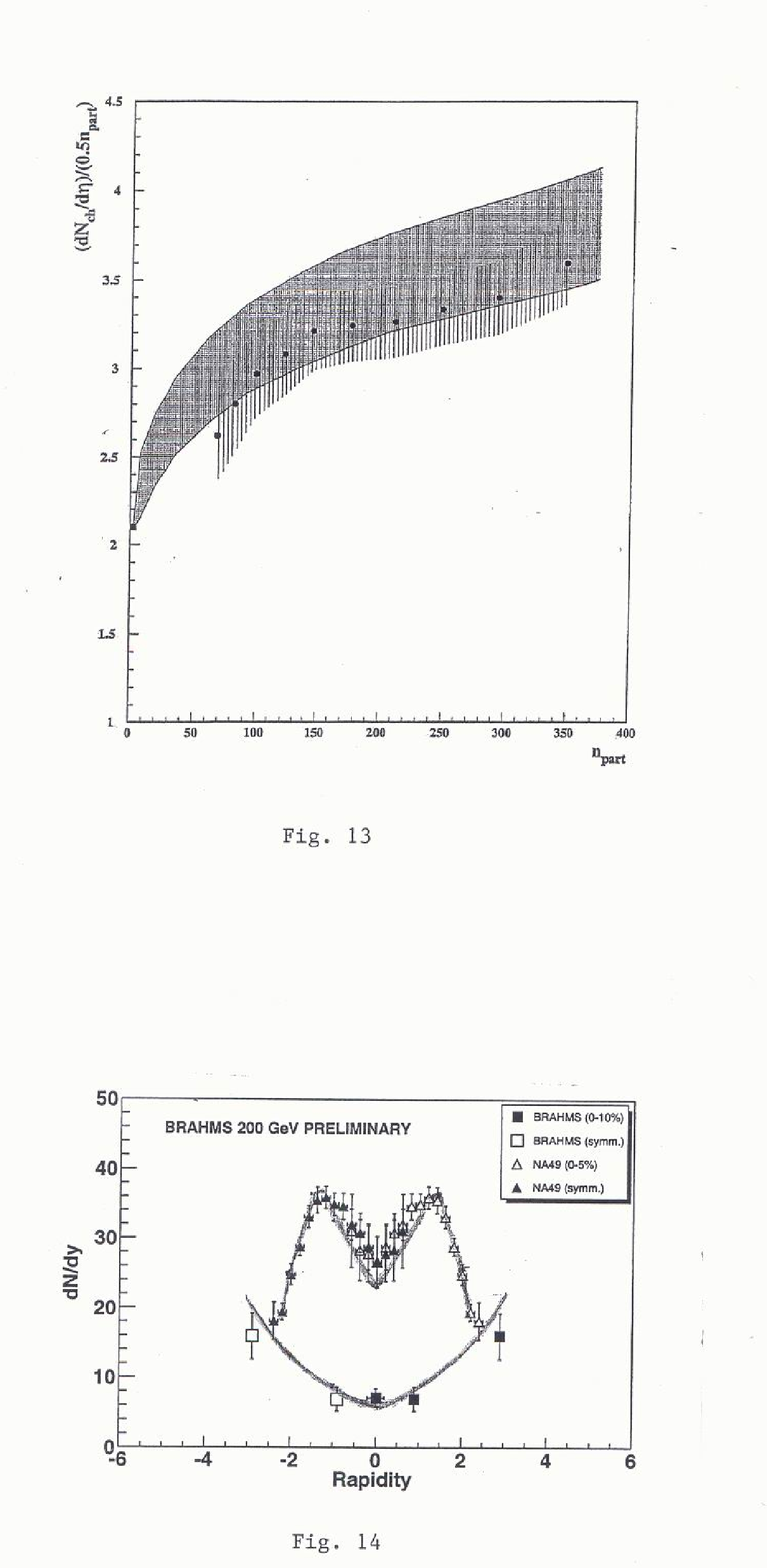}}}

\centerline{{\epsfxsize15cm \epsfbox{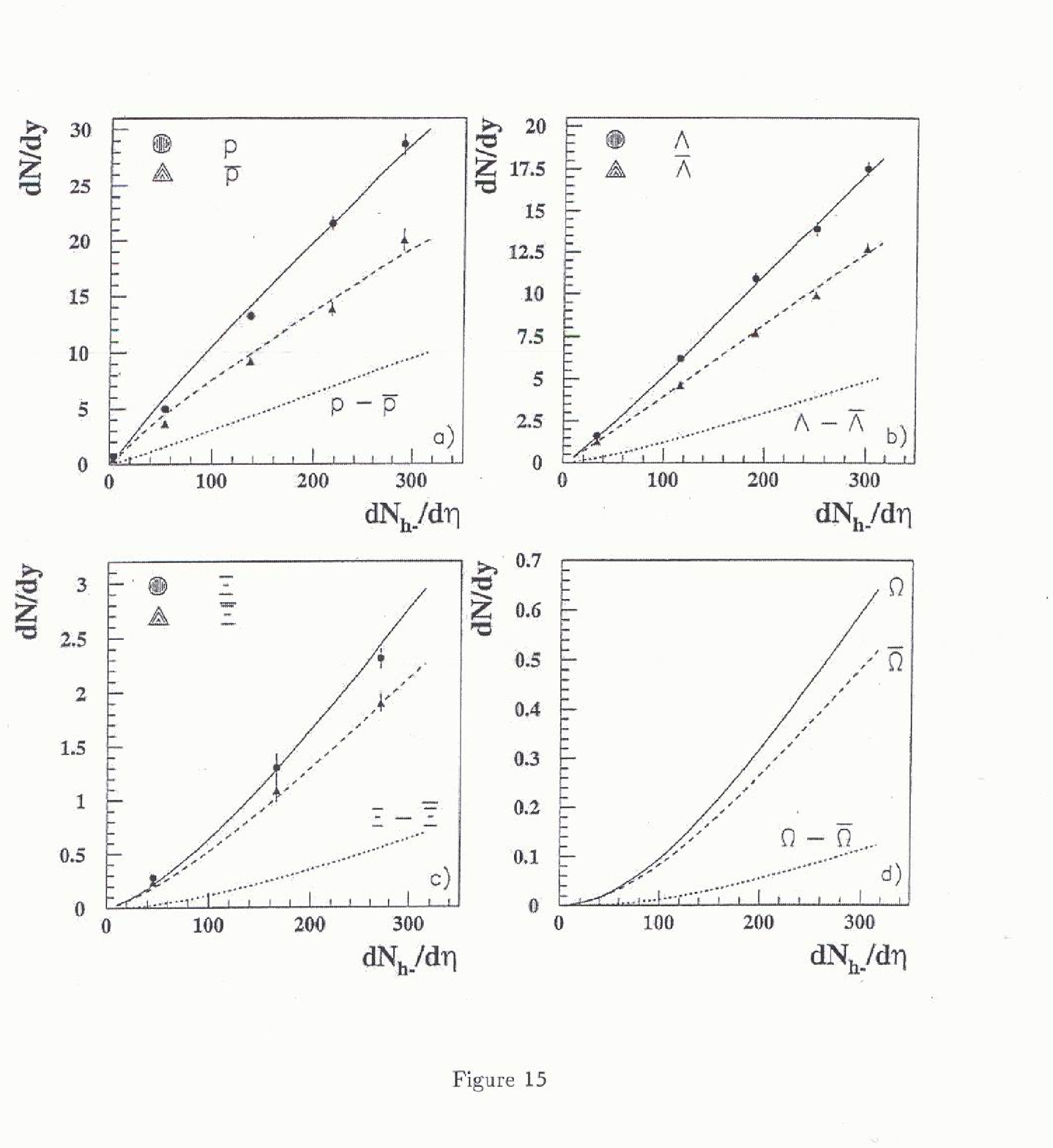}}}

\centerline{{\epsfxsize15cm \epsfbox{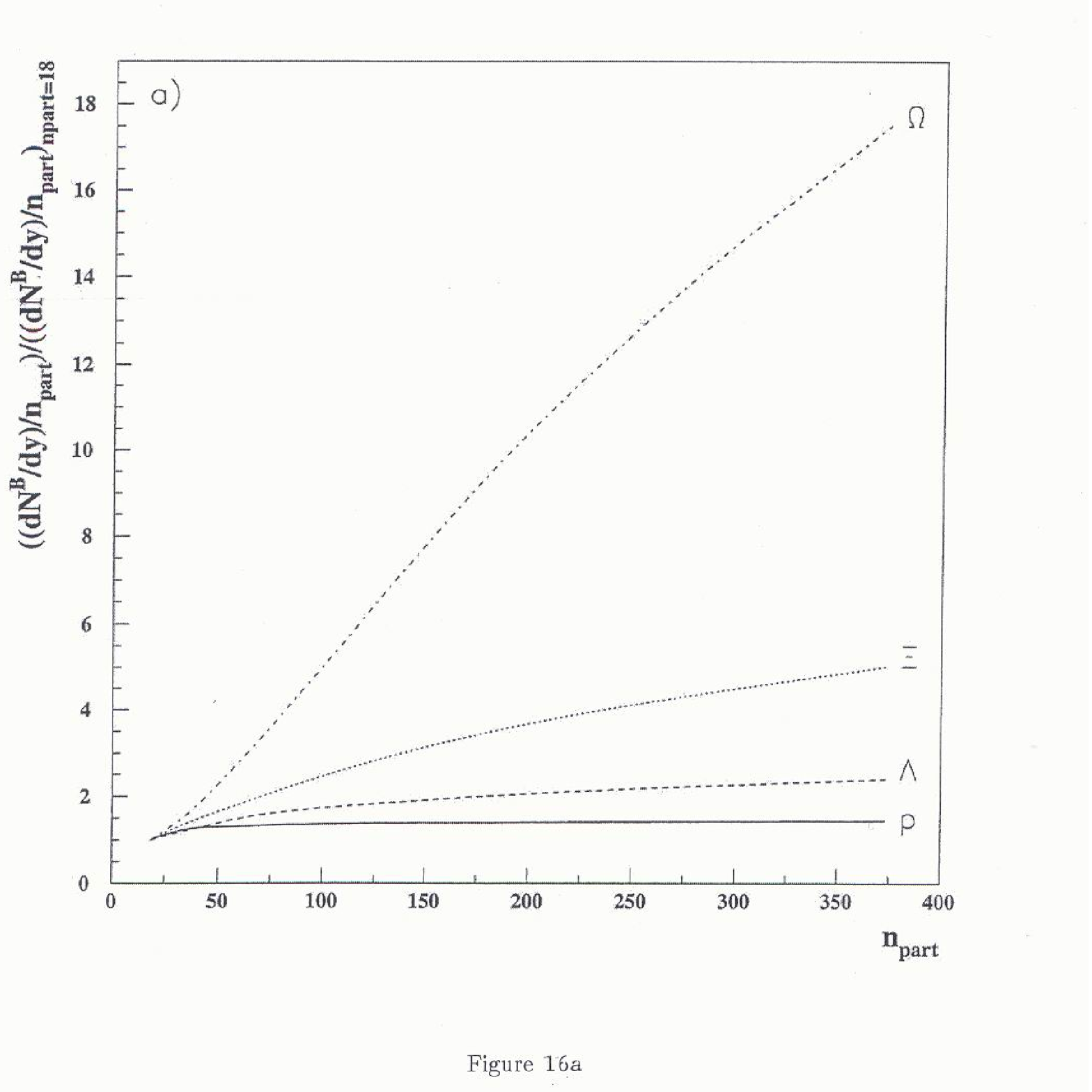}}}

\centerline{{\epsfxsize15cm \epsfbox{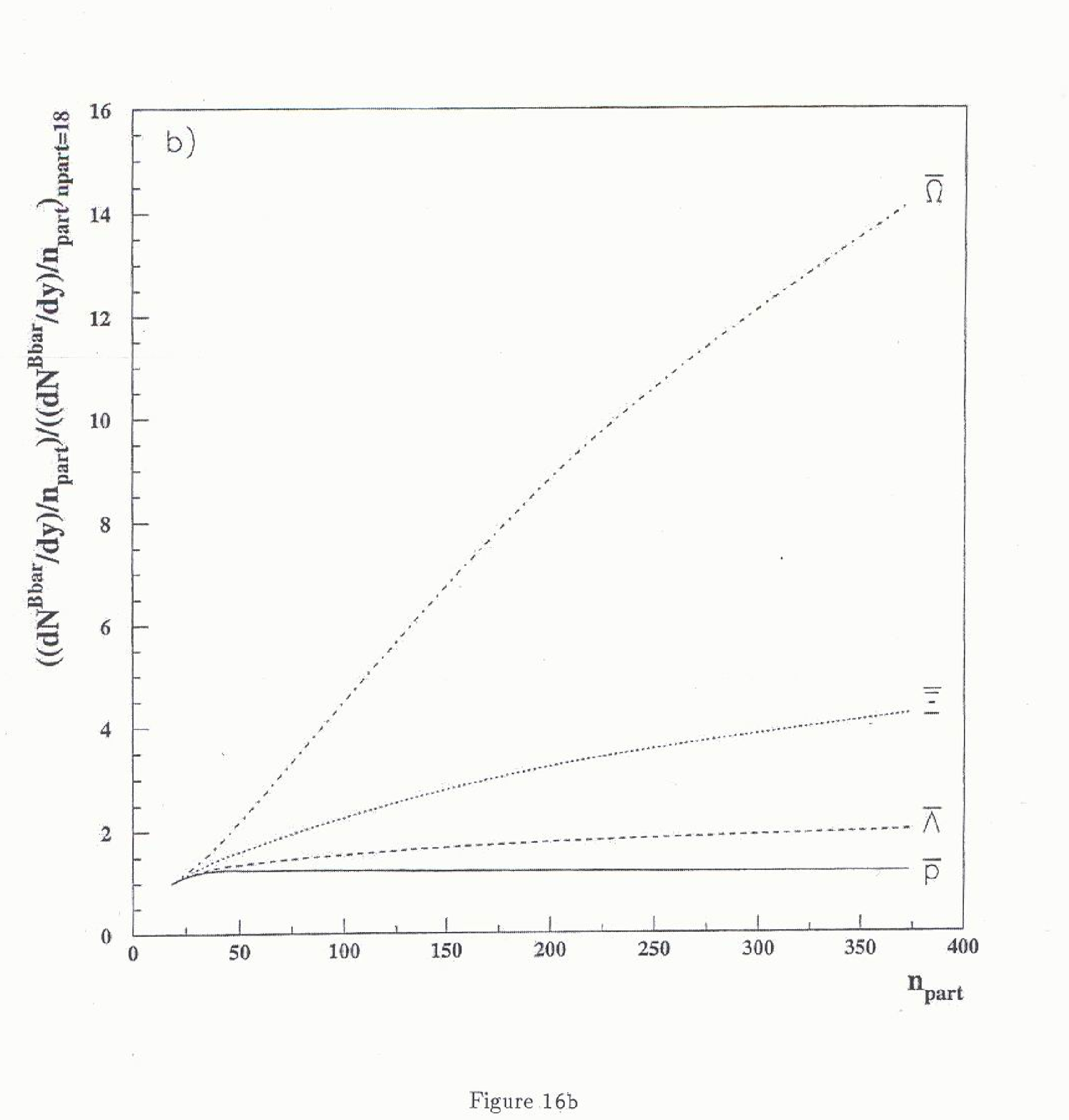}}}

\centerline{{\epsfysize21cm \epsfbox{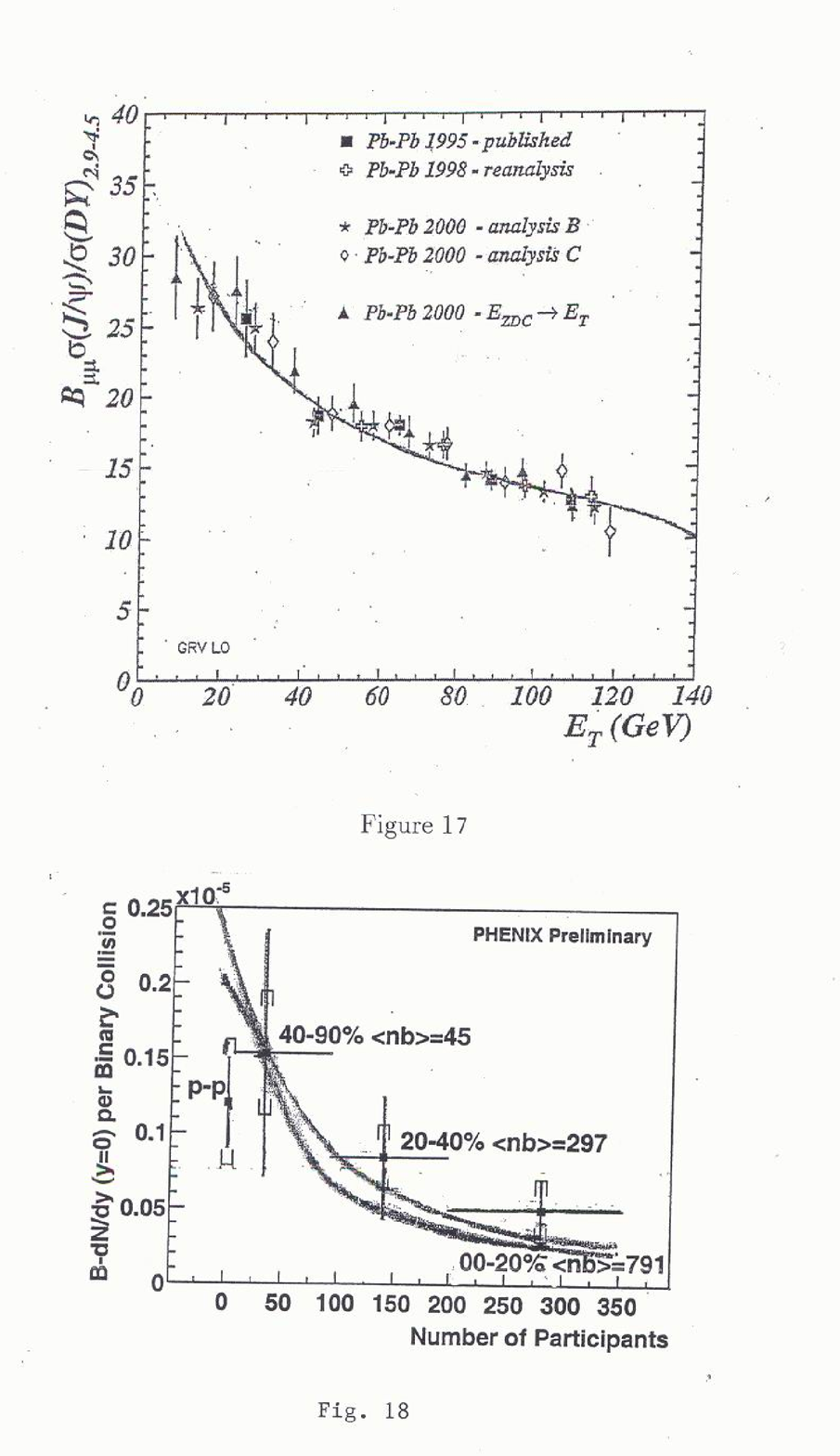}}}

\end{document}